\documentclass[fleqn,10pt]{wlscirep}
\usepackage[utf8]{inputenc}
\usepackage[T1]{fontenc}
\usepackage{bm} 
\usepackage{soul} 
\usepackage[table,xcdraw]{xcolor}

\title{Detecting gene-environment interactions to guide personalized intervention: boosting distributional regression for polygenic scores}

\author[1,*]{Qiong Wu}
\author[1,2]{Hannah Klinkhammer}
\author[3]{Kiran Kunwar}
\author[4,5]{Christian Staerk}
\author[3]{Carlo Maj}
\author[1]{Andreas Mayr}
\affil[1]{Institute for Medical Biometry and Statistics, Marburg University, Germany}
\affil[2]{Institute for Genomic Statistics and Bioinformatics, University of Bonn, Germany}
\affil[3]{Center for Human Genetics, Marburg University, Germany}
\affil[4]{IUF-Leibniz Research Institute for Environmental Medicine, Düsseldorf, Germany}
\affil[5]{Department of Statistics, TU Dortmund University, Germany}

\affil[*]{qiong.wu@uni-marburg.de}

\begin{abstract}
Polygenic risk scores can be used to model the individual genetic liability for human traits.  Current methods primarily focus on modeling the mean of a phenotype neglecting the variance. However, genetic variants associated with phenotypic variance can provide important insights to gene-environment interaction studies. To overcome this, we propose snpboostlss, a cyclical gradient boosting algorithm for a Gaussian location-scale model to jointly derive sparse polygenic models for both the mean and the variance of a quantitative phenotype. To improve computational efficiency on high-dimensional and large-scale genotype data (large $n$ and large $p$), we only consider a batch of most relevant variants in each boosting step. We  investigate the effect of statins therapy (the environmental factor) on low-density lipoprotein in the UK Biobank cohort using the new snpboostlss algorithm.  We are able to verify the interaction between statins usage and the polygenic risk scores for phenotypic variance in both cross sectional and longitudinal analyses. Particularly, following the spirit of target trial emulation, we observe that the treatment effect of statins is more substantial in people with higher polygenic risk scores for phenotypic variance, indicating gene-environment interaction. When applying to body mass index, the newly constructed polygenic risk scores for variance show significant interaction with physical activity and sedentary behavior. Therefore, the polygenic risk scores for phenotypic variance derived by snpboostlss have potential  to identify individuals that could benefit more from environmental changes (e.g. medical intervention and lifestyle changes).
\end{abstract}

\begin{document}

\flushbottom
\maketitle
%
%
\thispagestyle{empty}

\section*{Introduction}

Complex phenotypes are often influenced by various genetic and environmental factors as well as their interactions\cite{manolio2009finding}. Genome-wide association studies (GWAS) are able to detect many replicable genetic associations with various phenotypes\cite{yengo2018meta}. However, the endeavor to identify interactions between genetic variants and environmental factors (GxE) has so far achieved only limited success\cite{young2016multiple}. This may be because many traits are polygenic in nature, the effect sizes of GxE at individual variant level are often small, and a genome-wide scan leads to high multiple testing burden\cite{aschard2012challenges}. One alternative approach is to derive polygenic risk score (PRS) which measures the overall genetic predisposition for a phenotype and then to test for interactions between PRS and environmental factors\cite{belsky2018genetic,boyle2017expanded,fletcher2021health,schmitz2016long,barcellos2018education}. Typically, PRSs are computed as weighted sums of risk allele counts across genetic loci, with weights determined by GWAS-based summary statistics of univariate effects on the phenotypic mean. However, traditional PRS may not necessarily provide an accurate characterization of the genetic component in GxE interactions \cite{tang2022iprs}. 

\begin{figure}[ht]
\centering
\includegraphics[scale=0.7]{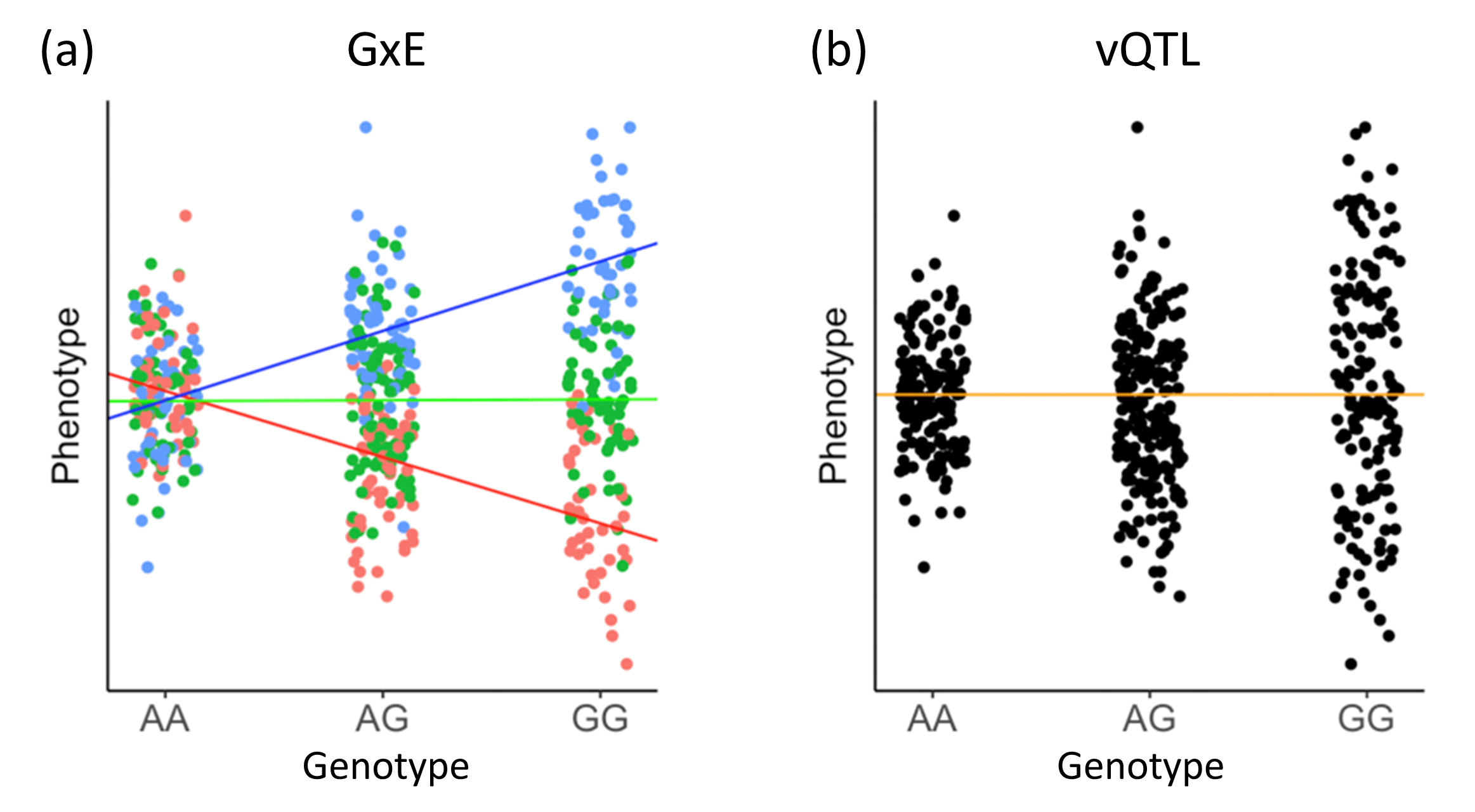}
\caption{Conceptual illustration showing that genetic variants in GxE affect the phenotypic variance with simulated data. (a) Different colors represent different levels of environmental factor. The effects of the genetic variant on the phenotype conditional on environmental levels are represented by the slopes of fitted lines. (b) Unconditional genetic effect on the phenotype, illustrated by the same data as (a). }
\label{fig:GxE_vQTL}
\end{figure}

Instead of using PRS derived from mean-regression models for GxE analyses, a more sensitive approach for detecting environmental effects is to prioritize variants associated with phenotypic variance (vQTLs) as candidates for G×E testing (Figure~\ref{fig:GxE_vQTL}). For a genetic variant which shows interaction with an environmental factor, its effect on the phenotype changes with environmental levels (Figure~\ref{fig:GxE_vQTL}(a)). However, as illustrated in Figure~\ref{fig:GxE_vQTL}(b), when we aggregate all environmental levels together, we can observe heteroscedasticity across genotype groups. In reverse, stratification of phenotypic variance gives rise to heteroscedasticity across genotype groups (Figure~\ref{fig:GxE_vQTL}(b)) and may reflect an underlying gene–environment interaction (Figure~\ref{fig:GxE_vQTL}(a)), as it indicates genotype-dependent modulation of phenotypic variability (Figure~\ref{fig:GxE_vQTL}(b)). Therefore, we can use the genetic variants associated with the phenotypic variance (variance quantitative trait loci [vQTLs]) as candidates to screen for GxE interactions\cite{marderstein2021leveraging, ronnegaard2011detecting,wang2019genotype, young2018identifying, miao2022quantile}. The idea of PRS, which predicts the mean of the continuous phenotype\cite{zhao2021pumas} or the risk for a disease, has also been extended to predict phenotypic variance by aggregating genetic effects across the whole genome, which is often referred to as variance polygenic risk score (vPRS)\cite{johnson2022polygenic,conley2018sibling, schmitz2016long}. The vPRS reflecting the genetic contribution to phenotypic plasticity, has gained recent successes in GxE analysis \cite{johnson2022polygenic,schmitz2016long}. But the currently available vPRS methods focus on estimating the phenotypic variability separately from the mean. 

We propose a method which can model both the mean and the variance simultaneously based on distributional regression. In this way, we do not only create an efficient way to derive polygenic models for both mean and variance, but also take into account the mutual influence between the two. What is simultaneously optimized is the likelihood function which incorporates both the predicted mean and the predicted variance.  Algorithm-wise, we built on the snpboost framework \cite{klinkhammer2023statistical, klinkhammer2024genetic} which applies adapted gradient boosting to select the most informative variants for mean prediction. The traditional boosting algorithm \cite{buhlmann2003boosting, buhlmann2007boosting} is adapted by adding a batch-building procedure so that each boosting iteration only works on a small batch of the most relevant variants. This can largely enhance computational efficiency and make it feasible to fit multivariable models on large-scale and high-dimensional genotype data (large $n$ and large $p$) as we typically encounter when developing PRS. Our proposed method, termed as snpboostlss, is an extension of snpboost into a distributional regression \cite{kneib2023rage} context, allowing us to implement variant selection and effect estimation and to construct PRSs for multiple distributional parameters simultaneously.

We demonstrate through simulation studies that the mPRS and vPRS derived from the proposed snpboostlss approach are efficient proxies for phenotypic mean and within-individual phenotypic variability, respectively. Afterwards, we apply snpboostlss on two phenotypes in UK Biobank\cite{bycroft2018uk} (UKBB): low density liproprotein (LDL) and body mass index (BMI). Both are considered to be subject to GxE interactions. When investigating LDL, we considered the use of statins as environmental factor. The
interaction between statins and vPRS is verified using both baseline and longitudinal data. We also mimicked a randomized controlled trial and found that the treatment effect of statins is more substantial in
people with higher vPRS, indicating gene-environment interaction. When applying to BMI, the constructed vPRS shows significant interaction with lifestyle variables such as physical activity and
sedentary behavior. Overall, our work highlights the advantage of the proposed snpboostlss approach in simultaneous and efficient modeling of phenotypic mean and variance for polygenic prediction and gene-environment interaction analysis.

\section*{Results}


\subsection*{Method overview}

For a quantitative phenotype $y_i$, we consider the  Gaussian location-scale model
\begin{equation}\label{eqn:gaussian_ls_res}
   \hspace{2in} y_i \overset{\mathrm{ind.}{}}{\sim} N(\mu_i, \sigma_i^2),\quad\mu_i=\bm{x}_i'\bm{\beta}, \quad \log(\sigma_i)=\bm{z}_i'\bm{\gamma}, \quad i = 1, \dots, n, 
\end{equation}
where location ($\mu_i$) and scale with log-link ($\log(\sigma_i)$) are modeled as aggregate linear effects of selective informative genome-wide bi-allelic single-nucleotide polymorphisms (SNPs) contained in $\bm{x}_i$ and $\bm{z}_i$, respectively.  The vectors $\bm{x}_i$ and $\bm{z}_i$ can represent different subsets of variants. The goal of our proposed snpboostlss algorithm is to identify the informative $\bm{x}_i$ and $\bm{z}_i$ from genome-wide genotype data and simultaneously estimate their effects $\bm{\beta}$ and $\bm{\gamma}$.

This goal is achieved by component-wise gradient boosting for distributional regression\cite{mayr2012generalized} with the likelihood representing the objective function. To overcome the computational issue due to large scale and high dimensionality of the genotype data, we implemented a batch-building procedure on top of the boosting process so that each boosting iteration only works on a small subset of most relevant variants\cite{klinkhammer2023statistical}. Apart from a training set for boosting, we also utilized a separate validation set to determine the stopping iteration as the main tuning parameter. This is possible given the large sample size usually available in databases such as UK Biobank. This approach avoids computationally heavy tuning methods such as cross-validation. Given two sets of selected variants and their estimated effect sizes from the algorithm, we can further construct, for each individual, two polygenic risk scores $\text{mPRS}_i:=\bm{x}_i'\bm{\hat\beta}$ and $\text{vPRS}_i:=\bm{z}_i'\bm{\hat\gamma}$. More details on the algorithm can be found in \emph{Methods} and in the supplementary information (\emph{SI}, Section S1).

\subsection*{Simulation results}

We conducted a simulation study to investigate the performance of the new approach under known conditions, looking particularly at these two specific aims: (i) to compare mPRS with that derived by the established snpboost algorithm by Klinkhammer et al.\cite{klinkhammer2023statistical}, and (ii) to compare vPRS with within-individual variability estimator using longitudinal data. 
 
Simulations were based on HAPNEST synthetic genotype data \cite{wharrie_hapnest_2023} which preserve the key properties of large-scale biobank databases. Continuous phenotypes were generated from the Gaussian location-scale model with genetically driven mean and variance. To account for different genetic architectures, we considered varying heritability $h^2$ and sparsity $s$, defined as the proportion of total phenotypic variance explained by mPRS and the proportion of informative variants, respectively. Simulated datasets were randomly split into 50\% training, 20\% validation and 30\% test sets. We used training and validation sets for model fitting and test set for performance evaluation. See \emph{Methods} for detailed description on simulation settings.

\begin{figure}[ht]
\centering
\includegraphics[width=\linewidth]{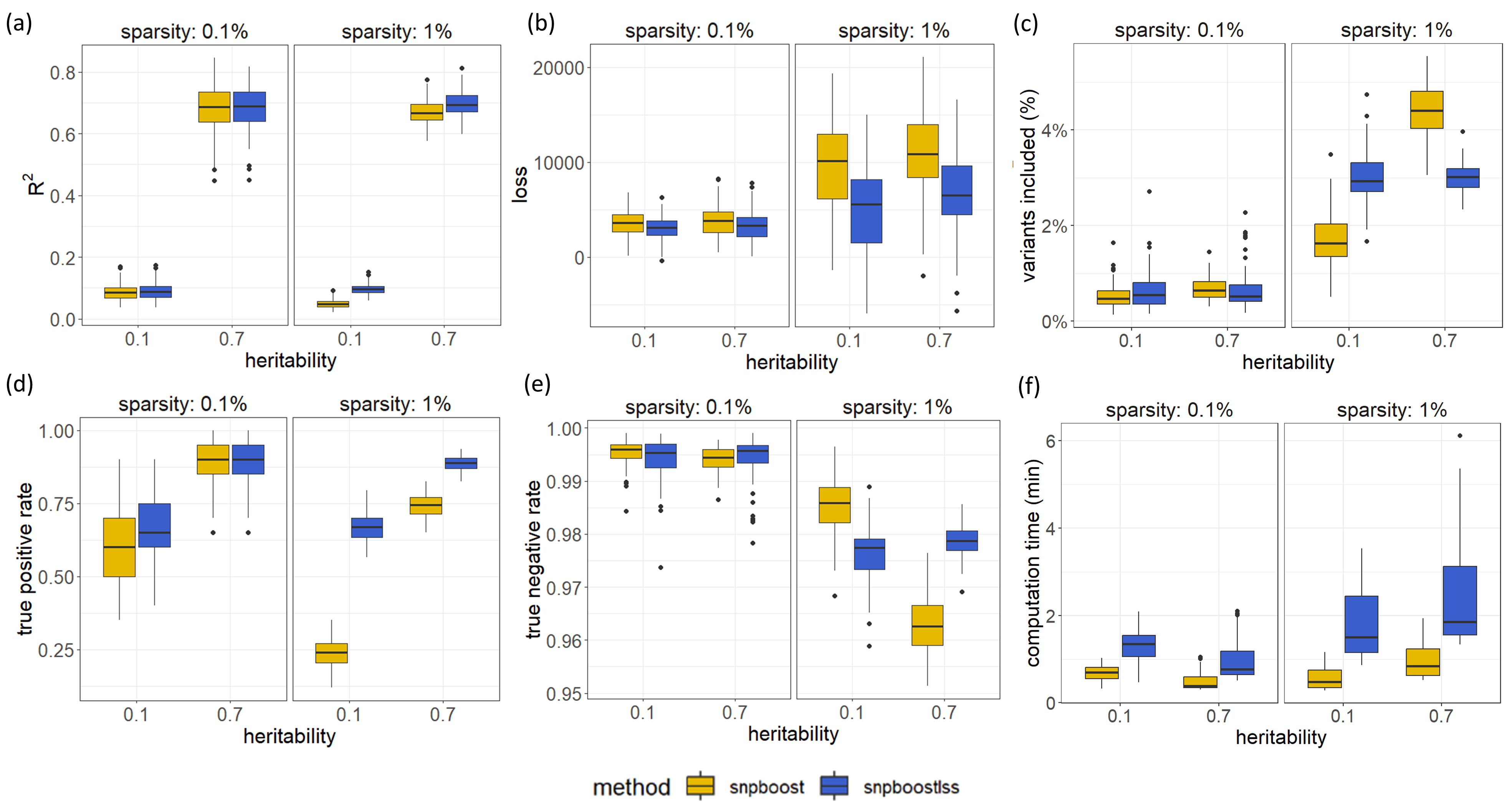}
\caption{Comparison between snpboostlss and snpboost on modeling mPRS (location parameter). Results of scenarios with heritability $h^2\in\{0.1,0.7\}$ and sparsity $s\in\{0.1\%,1\%\}$ for $p=20,000$ variants and $n=20,000$ individuals (divided into 50\% training, 20\% validation and 30\% test sets) are shown. For each performance metric, the boxplots from 100 simulations are displayed.}
\label{fig:sim_res_aim2_main}
\end{figure}

We first compared snpboostlss with snpboost with focuses on the prediction performance and variant selection for mPRS. We investigated whether modeling additionally the phenotypic variance (vPRS) could impact the performance in estimating the mean (mPRS) when there exists heteroscedasticity. Figure~\ref{fig:sim_res_aim2_main}(a) indicates that snpboostlss can capture the true heritability more accurately, i.e., the $R^2$ achieved from snpboostlss is closer to true heritability (0.1 or 0.7). As shown in Figure~\ref{fig:sim_res_aim2_main}(b), snpboostlss yields lower loss defined as negative log-likelihood, especially when sparsity level is 1\%.  Figure~\ref{fig:sim_res_aim2_main}(c) shows that, given certain sparsity level snpboostlss selects similar number of variants regardless of heritability, while snpboost tends to select more when the effect sizes of informative variants are larger. However, both methods tend to overestimate the number of informative variants, which is a common characteristic of boosting algorithms\cite{stromer2022deselection}. In addition, snpboostlss exhibits superior variable selection accuracy in terms of true positive rate, as manifested in Figure~\ref{fig:sim_res_aim2_main}(d). The average true positive rates of snpboostlss are almost 0.9 when heritability is 0.7, indicating that a majority of the informative variants are correctly identified in these scenarios. Even when the signals are weak (heritability = 0.1), more than 60\% of informative variants can still be selected by snpboostlss. Figure~\ref{fig:sim_res_aim2_main}(e) demonstrates that given the sparsity level, snpboostlss tends to get a similar true negative rate regardless of heritability, but snpboost yields higher true negative rates for lower heritability. This corresponds to Figure~\ref{fig:sim_res_aim2_main}(c) where snpboostlss selects similar number of variants regardless of heritability, while snpboost is more conservative when heritability is low. Finally, as shown in Figure~\ref{fig:sim_res_aim2_main}(f), snpboostlss requires longer computation time than snpboost. This is expected because it models both mPRS and vPRS and the algorithm hence needs to circle through roughly twice as many base-learners. To summarize, in our simulations where individual phenotypic variance can differ, modeling additionally the phenotypic variance (vPRS) could improve the performance in estimating the mean (mPRS) in terms of prediction and informative variants detection. This advantage is more notable when there are more informative variants with larger effect sizes. It's worth noting, however, that such advantage may be partially due to the concordance between simulation setting and distributional assumption of snpboostlss. Even in such situations, mean regression approaches like snpboost\cite{klinkhammer2023statistical, klinkhammer2024genetic} already provide good phenotype prediction performance in various scenarios.

\begin{figure}[!ht]
\centering
\includegraphics[width=\linewidth]{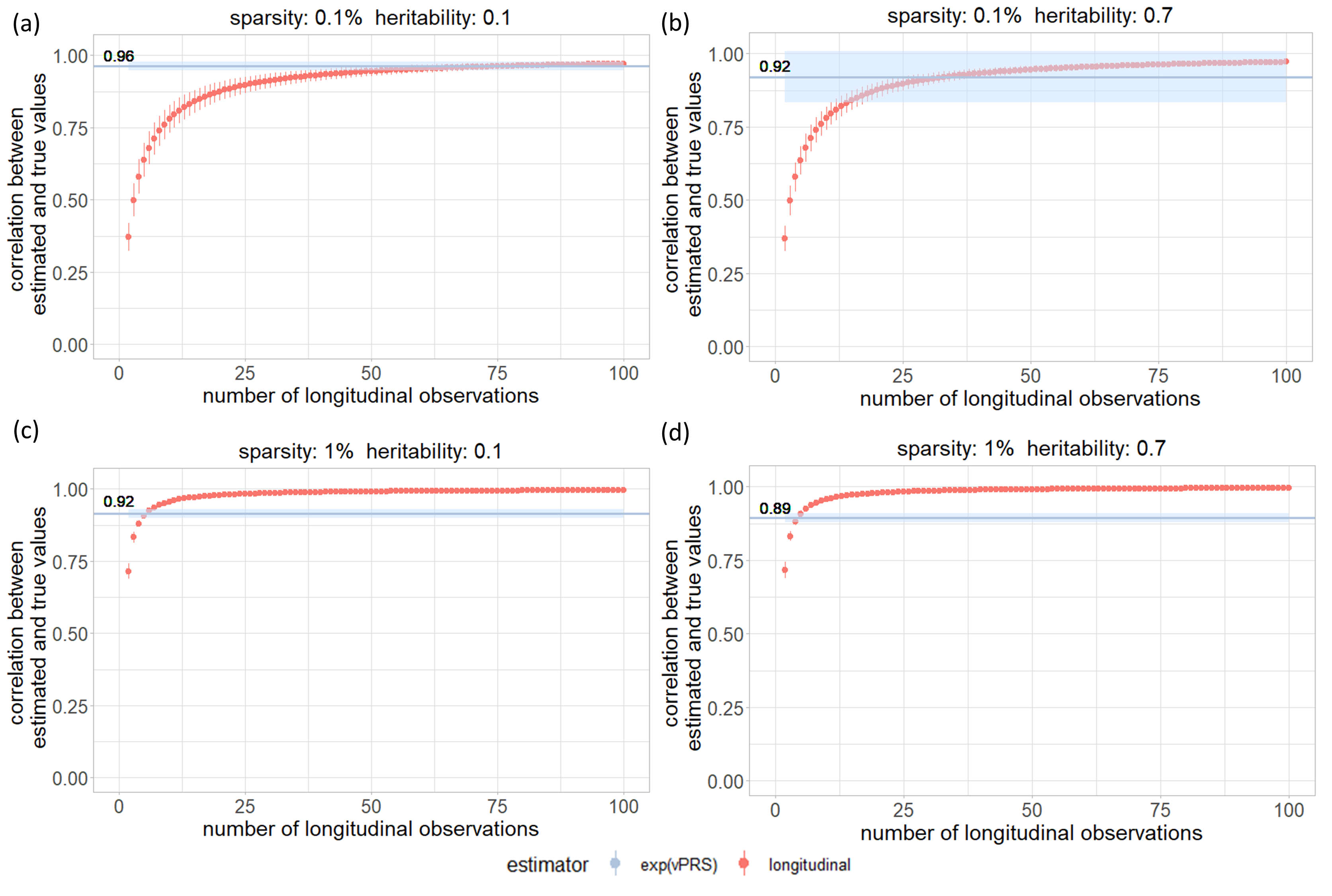}
\caption{Comparison between vPRS estimator and within-individual sample SDs calculated using longitudinal data as naive benchmark on the accuracy of estimating the within-individual variability $\sigma_i$ (scale parameter). Results of scenarios with heritability $h^2\in\{0.1,0.7\}$ and sparsity $s\in\{0.1\%,1\%\}$  are shown. $corr(\sigma_i, \hat{\sigma_i})$ was calculated on the test set with 6,000 subjects. For each performance metric, the mean$\pm$SD from 100 simulations are displayed.}
\label{fig:sim_res_aim3_main}
\end{figure}

Secondly, we evaluated the accuracy of vPRS derived by snpboostlss using baseline data on estimating within-individual phenotypic variability. With the simulated longitudinal data, we can obtain a naive benchmark estimator for within-individual phenotypic variability; that is the standard deviation (SD) of each individual's repeated phenotype measurements (without taking genetic information into account). The within-individual sample SDs were calculated using 2, 3, ..., 100 repeated measurements, respectively. In most practical settings these numbers of repeated measurements will not be available, but in this artificial simulation scenario the sample SDs can serve as a natural benchmark. The accuracy of these estimators is assessed by the correlation between estimated and true $\sigma_i$'s. Figure~\ref{fig:sim_res_aim3_main}(a) shows that vPRS estimator is as good as benchmark estimator constructed with approximately 70 longitudinal observations when 0.1\% variants are informative and heritability is 0.1. If heritability increases from 0.1 to 0.7 (Figure~\ref{fig:sim_res_aim3_main}(b)), the proportion of phenotypic variance that cannot be explained by mPRS decreases, which actually makes it harder to detect informative variants for $\sigma$. This leads to a reduced average correlation from 0.96 to 0.92. However, in this case, the vPRS estimator is still comparable to the benchmark estimator using approximately 35 longitudinal observations. When the proportion of informative variants increases from 0.1\% to 1\% (Figure~\ref{fig:sim_res_aim3_main}(c) and \ref{fig:sim_res_aim3_main}(d)), our vPRS can still retain a correlation around 0.9 regardless of heritability levels. It is also interesting to notice that the performance of the benchmark estimator improves greatly when the proportion of informative variants increases. The main reason is an increase in the variance of our generated $\sigma_i$ across individuals when there are more informative variants. Part of this variance that cannot be explained by the variance of the benchmark estimator is characterized by the estimation error of benchmark estimator, which would stay at similar magnitude given the number of longitudinal observations used for estimation. Therefore, the unexplained proportion drops and the explained proportion rises correspondingly, leading to higher correlation between longitudinal estimators and true values of $\sigma_i$. As a consequence, the benchmark estimator with only three to five longitudinal observations can match the estimation accuracy of vPRS. In summary, our vPRS using genotype information and only the baseline phenotype can provide accurate estimation for within-individual variability. Comparing with the naive benchmark estimator derived with longitudinal data, the vPRS shows favorable estimation accuracy when the longitudinal observations are not abundant, thus providing an efficient proxy for within-individual phenotypic variability.

\subsection*{Identification of variants in mPRS and vPRS for LDL in UK Biobank}

We applied snpboostlss on the LDL data of unrelated subjects with British ancestry in the UK Biobank. After quality control, 244,583 individuals with genotype data containing 604,967 bi-allelic SNPs on autosomes and LDL measurements were included in the analysis (\emph{Methods}). These subjects were split into training, validation and test sets with allocation ratio of 2:1:1. We trained mPRS and vPRS models by running snpboostlss on training and validation sets, then investigated GxE interactions on the test set. The distribution of the LDL shows slight right skewness (\emph{SI}, Figure S5), therefore the Gaussian location-scale model in Equation \eqref{eqn:gaussian_ls_res} is a reasonable approximation. Running snpboostlss on a high performance cluster with 2 CPUs and 12 GB memory per CPU took around 16 minutes.  

\begin{figure}
    \centering
    \includegraphics[scale=0.8]{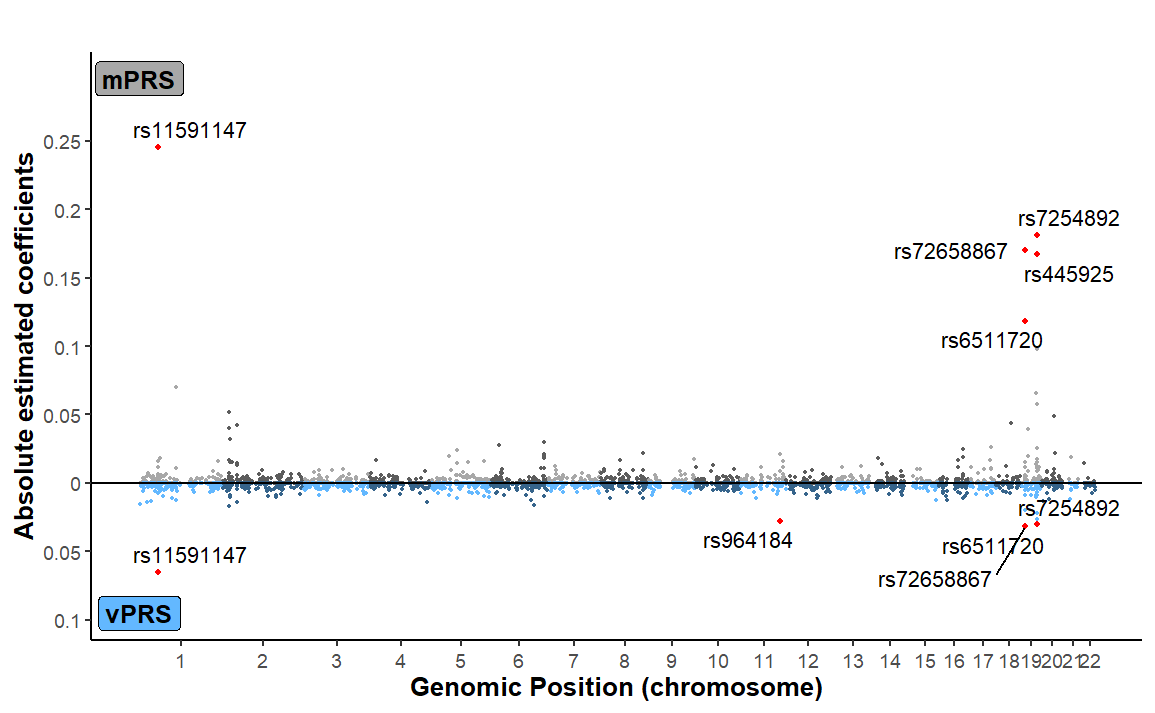}
    \caption{Absolute estimated effect sizes of variants in mPRS and vPRS, fitted by snpboostlss on UK Biobank data with LDL as phenotype. Variants are ordered based on their location at the genome. Variants with five largest absolute coefficient size in mPRS and vPRS are annotated.}
    \label{fig:manhatten_LDL}
\end{figure}

The resulting snpboostlss model includes 713 variants in mPRS and 979 variants in vPRS with 58 variants shared by both, meaning that they affect both mean and variance of LDL. Mapping all selected variants to linkage disequilibrium blocks (LD blocks) reveals a total of 889 LDL-associated LD blocks (466 for mPRS and 660 for vPRS). 26.7\% of these LD blocks (237) are shared between mPRS and vPRS, showing a higher degree of overlap at the LD-block resolution than at the genetic-variant resolution. We visualized the effect size and genome position of selected variants in Figure~\ref{fig:manhatten_LDL}. Most of the leading variants in mPRS and vPRS come from the same regions of the genome. We looked into more details about the top five variants with largest absolute effect size in mPRS and vPRS (Table~\ref{tab:top_variants_LDL}). Four of them are the same and the rest (rs445925 for mPRS and rs964184 for vPRS) are also shared variants for mPRS and vPRS. These top variants have all been considered as LDL-associated in the existing literature \cite{de2019multiancestry,bentley2019multi,loya2025scalable,zhu2017susceptibility}, and are mapped to genes well-known to be associated with LDL such as \emph{PCSK9}, \emph{NECTIN2}, \emph{LDLR} and \emph{ZPR1}. An additional gene annotation enrichment analysis of the vPRS gene set associated with the LDL cholesterol revealed a strong enrichment on terms such as LDL levels ($P= 3.328\times 10^{-30}$ , $OR$ = 6.85), total cholesterol levels ($P= 2.550\times 10^{-22}$ , $OR$ = 9.57) and medication used to lower cholesterol levels in blood (Hmg Coa Reductase Inhibitors, commonly known as statins; $P= 2.159\times 10^{-18}$ , $OR$ = 12.33).

\begin{table}
\caption{Top five variants in mPRS and vPRS selected by snpboostlss for LDL. Their rsID, mapped genes (GRCh37/hg37) and association to LDL in existing literature are reported.}
\label{tab:top_variants_LDL}
\begin{tabular}{cccc|cccc}
\toprule
\multicolumn{4}{c|}{mPRS}   & \multicolumn{4}{c}{vPRS}     \\ \hline
Rank & SNP        & Gene              & LDL-related & Rank & SNP         & Gene              & LDL-related \\ \hline
1    & rs11591147 & \emph{PCSK9}      & Yes\cite{de2019multiancestry}         & 1    & rs11591147  & \emph{PCSK9}      & Yes\cite{de2019multiancestry}         \\
2    & rs7254892 & \emph{NECTIN2}      & Yes\cite{bentley2019multi}         & 2    & rs6511720  & \emph{LDLR}       & Yes\cite{bentley2019multi}         \\
3    & rs72658867 & \emph{LDLR}      & Yes\cite{loya2025scalable}         & 3    & rs72658867  & \emph{LDLR}      & Yes\cite{loya2025scalable}         \\
4    & rs445925  & \emph{APOE, APOC1} & Yes\cite{zhu2017susceptibility}         & 4    & rs7254892   & \emph{NECTIN2} & Yes\cite{bentley2019multi}         \\
5    & rs6511720  & \emph{LDLR}              & Yes\cite{bentley2019multi}         & 5    & rs964184 & \emph{ZPR1}      & Yes\cite{bentley2019multi}          \\ \bottomrule
\end{tabular}
\end{table}

\subsection*{Detection of GxE for LDL using baseline data}

We investigated whether the variants in vPRS are involved in GxE interactions for LDL. This was carried out by testing whether the constructed vPRS can show interaction effects with environmental factors. The environmental factor considered here is the use  of any statins which are commonly prescribed medications to lower LDL\cite{kapur2008clinical}. We considered statins usage as a binary variable and investigated its main effect and interaction effect with vPRS through the following model:
$$\text{LDL}_i \sim \text{mPRS}_i + \text{vPRS}_i + \text{statins}_i + \text{vPRS}_i\times \text{statins}_i$$
where mPRS is to adjust for predicted average LDL level, vPRS is standardized (i.e., with zero mean and variance of one), LDL and statins usage are baseline observations of 61,145 subjects in the test set. We further adjusted for additional covariates (\emph{Methods}). The effect of statins in lowering LDL is verified by its negative main effect ($P< 2\times 10^{-16}$ , \emph{SI}, Table S1). More interestingly, the vPRS-statins interaction is also significantly negative ($P< 2\times 10^{-16}$, \emph{SI}, Table S1). That means that the total effect of statins on subjects with higher vPRS is more profound  (Figure~\ref{fig:LDL_application}(a)), so statins therapy can be more effective for them in lowering LDL. The interaction remains significant after we adjusted for additional vPRS-covariate interaction terms in the model ($P< 2\times 10^{-16}$, \emph{SI}, Table S1), indicating the robustness of our result.

In summary, we verified, with baseline data, that our constructed vPRS for LDL can show significant interaction with a relevant environmental factor, the use of statins, in the UK Biobank. Next, we would further verify such GxE interactions using longitudinal data, to investigate whether people with higher vPRS could indeed experience larger decrease in LDL after using statins. This was performed using longitudinal observations from UK Biobank in a self-controlled design and a parallel group design. 

\subsection*{Effect of statins to lower LDL in different vPRS groups: a self-controlled design}

In a self-controlled design, we filtered, in the test set, for those subjects who did not use statins at baseline but were using statins at the first revisit and had LDL measured at both visits. 767 subjects were eligible after filtering (\emph{SI}, Figure S4). We then measured the effect of statins therapy by calculating the change from baseline in LDL. We compared the statins effect between high-vPRS and low-vPRS groups, which are defined as the groups of people whose vPRS belong to either top/bottom quartile or top/bottom decile defined on the test set (\emph{Methods}). In Figure~\ref{fig:LDL_application}(b), the changes in LDL for both high-vPRS and low-vPRS groups are negative on average, and high-vPRS group shows significantly larger LDL drop than the low-vPRS group. In other words, the effect of statins in lowering LDL is more prominent for the people with higher vPRS, which verifies the GxE interaction observed from the baseline data. 

\subsection*{Effect of statins to lower LDL in different vPRS groups: a parallel group design}

To further strengthen our verification of GxE interaction, we considered a parallel group design with two treatment groups to mimick a randomized controlled trial (RCT) with more subjects included. In the test set, we filtered for people with baseline LDL higher than 3.36 mmol/L (130 mg/dl), which is a commonly used eligibility criteria in trials with statins as primary prevention of cardiovascular diseases \cite{adams2014rosuvastatin, zhao2018efficacy, talavera2013double, florentin2013colesevelam, her2010effects} (see \emph{Methods} for more discussion on the eligibility criteria). Then we included those subjects who did not take statins at baseline and whose LDL measurements and statins usage status at both baseline and first revisit are available. In the end, 1,276 subjects were included in the analysis set with 530 taking statins at first revisit (considered as intervention group) and 746 not on statins at first revisit (considered as control group) (\emph{SI} Figure S4). To analyze treatment effects with observational data in a parallel group design, we followed the spirit of target trial emulation \cite{hernan2022target} and performed inverse probability of treatment weighting (IPTW) \cite{chesnaye2022introduction} to adjust for potential confounding such that the confounders are equally distributed across two treatment groups. Details of IPTW can be found in \emph{Methods}.  

\begin{figure}
    \centering
    \includegraphics[height=0.8\textheight]{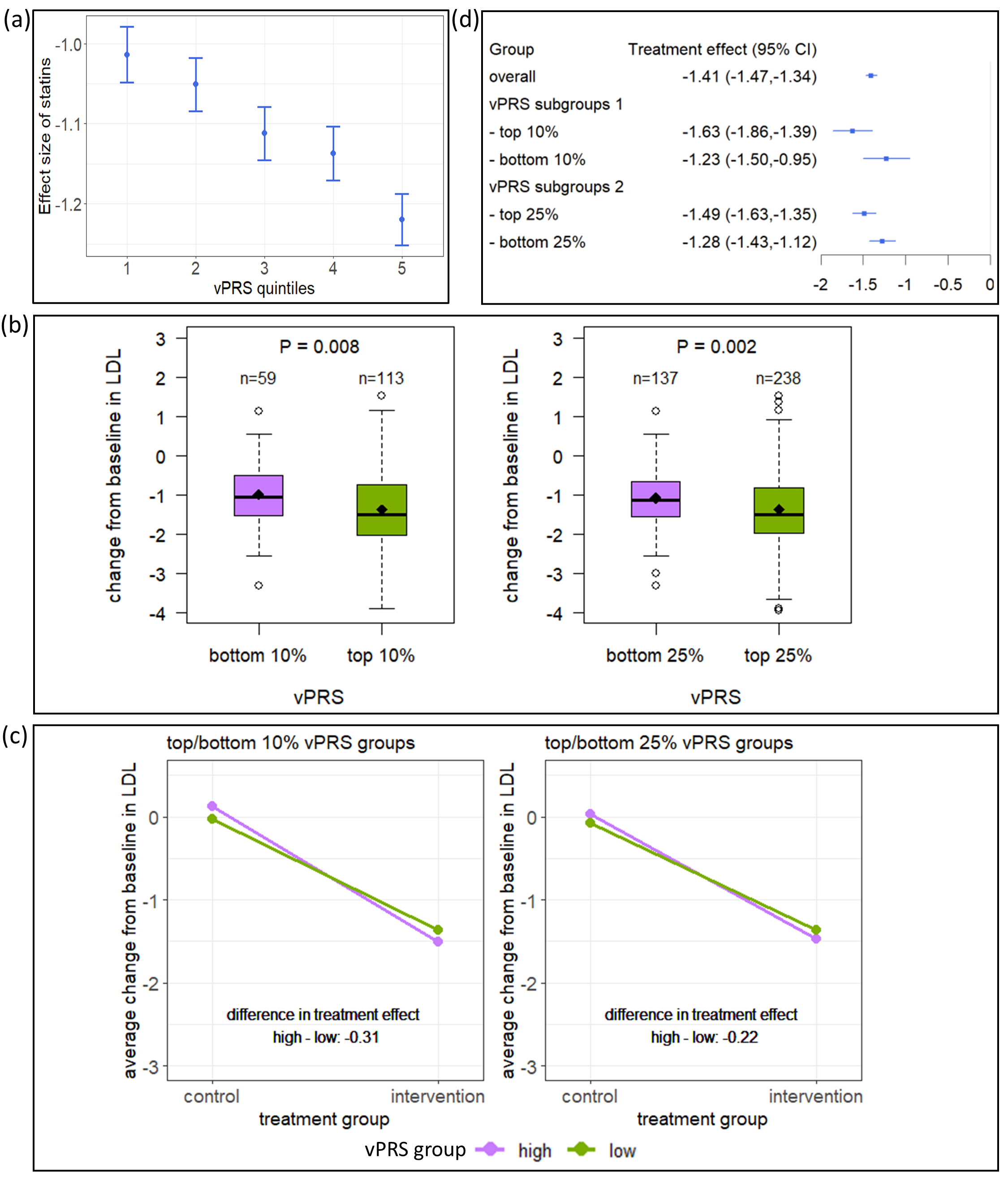}
    \caption{Verification of interaction between vPRS and statins usage status for LDL in UKBB. (a) Illustration of GxE on baseline data in the test set. For each quintile of vPRS, the estimated effect of statins on LDL along with 95\% CI is displayed. (b) Comparison of LDL change (mmol/L) between high-vPRS and low-vPRS groups in self-controlled design. The vPRS subgroups are defined as people with vPRS beyond 90\%/10\% percentile (left panel) and 75\%/25\% percentile (right panel) of vPRS in the test set. (c) Comparison of treatment effect of statins between high-vPRS and low-vPRS groups in parallel group design. In each plot, the four points represent the weighted average of LDL change from baseline (mmol/L) for the corresponding vPRS- treatment-subgroup. The weights are derived by IPTW. The slope of each line represents the treatment effect in the corresponding vPRS-subgroup, and the difference between the slopes of two lines represents the interaction effect between vPRS and statins. (d) Overall and vPRS-subgroup treatment effect of statins in parallel group design. Treatment effect is obtained from linear model with LDL change from baseline as response and adjusted for treatment group and other baseline covariates.}
    \label{fig:LDL_application}
\end{figure}

We considered the same vPRS-based subgrouping approaches as in the self-controlled design. Figure~\ref{fig:LDL_application}(c) illustrates that the treatment effect, measured by the difference of average change from baseline in LDL between intervention and control groups, is larger for the high vPRS group than that for the low vPRS group. Such tendency is more prominent when the subgroups are based on more extreme vPRS quantiles.

We further quantified the treatment effect in overall analysis set for the parallel group design including 1,276 subjects and in different vPRS subgroups using regression models (\emph{Methods}). The effect of statins to lower LDL is illustrated by the negative overall treatment effect and subgroup treatment effects (Figure~\ref{fig:LDL_application}(d)). In addition, the high-vPRS group experiences a larger treatment effect than the low-vPRS group, which is consistent with the observations in Figure~\ref{fig:LDL_application}(c). To additionally investigate whether such difference is significant, we performed the subgroup interaction analysis (\emph{Methods}), which is a usual part of subgroup analysis in RCTs. The interaction is found to be significant for both top/bottom 25\% vPRS subgrouping ($P= 1.21\times 10^{-3}$ ) and top/bottom 10\% vPRS subgrouping ($P = 1.44\times 10^{-2}$), which further verifies the GxE we detected using baseline data only. 

To summarize our analyses on LDL, we constructed mPRS and vPRS with our newly proposed snpboostlss algorithm on the UK Biobank data. Given the motivation demonstrated in Figure~\ref{fig:GxE_vQTL}, we formed the hypothesis that vPRS might serve as a proxy for the genetic component in GxE. We then verified our hypothesis through multiple sources of evidence with various data structures and study designs. Our results indicate that a potential use of the snpboostlss is to provide clinicians a tool to screen for people who can benefit more from environmental changes or even medical interventions (like statins) based on their vPRS.

\subsection*{Verification of GxE for BMI in UK Biobank}

We also considered another phenotype, BMI, to investigate whether vPRS can also work as an indication for the sensitivity to lifestyle changes. We utilized the observations at the initial visit in UKBB. After quality control, 351,891 individuals with genotype data containing 510,061 bi-allelic SNPs were included in the analysis (\emph{Methods}). BMI of these subjects exhibits slight right skewness in distribution (\emph{SI}, Figure S8). Running snpboostlss on a high performance cluster with 2 CPUs and 12 GB memory per CPU took around 30 minutes.  

\begin{figure}
    \centering
    \includegraphics[scale=0.8]{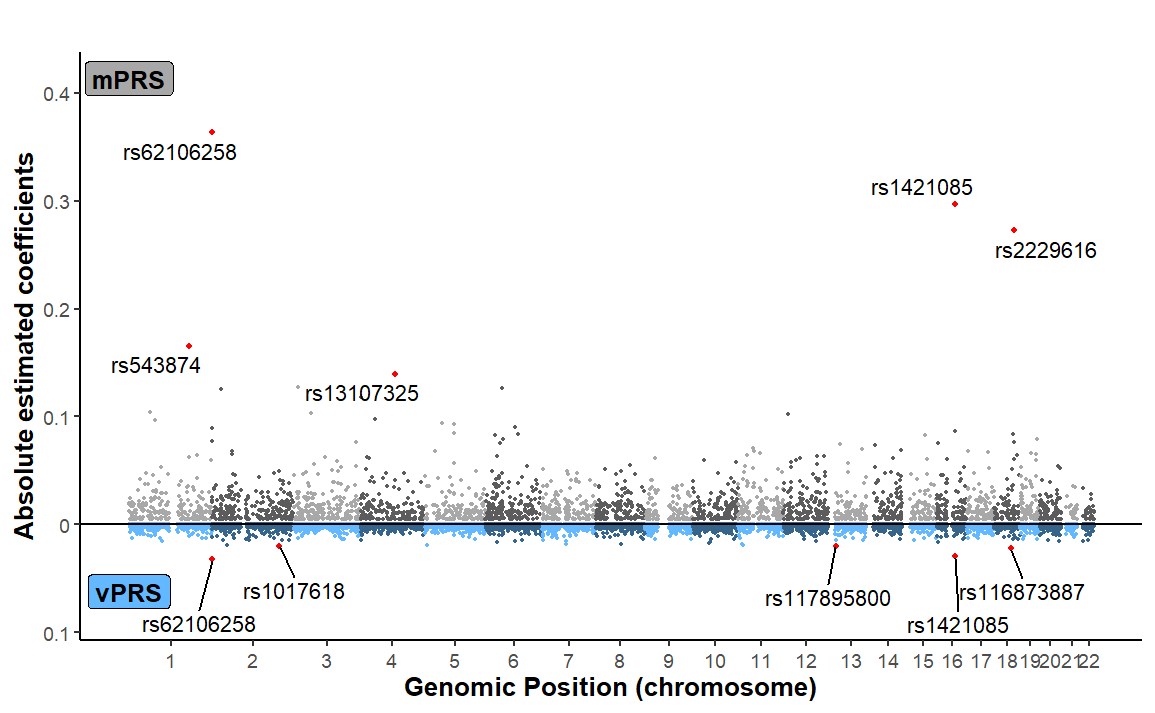}
    \caption{Absolute estimated effect sizes of variants in mPRS and vPRS, fitted by snpboostlss on UK Biobank data with BMI as phenotype. Variants are ordered based on their location at the genome. Variants with five largest absolute coefficient size in mPRS and vPRS are annotated.}
    \label{fig:manhatten_BMI}
\end{figure}

The resulting snpboostlss model includes 2,748 variants in mPRS and 3,430 variants in vPRS, between which 286 are shared variants. The selected variants are mapped to a total of 1,532 BMI-associated LD blocks (1,164 for mPRS and 1,365 for vPRS). The majority of these LD blocks (997) are shared between mPRS and vPRS, showing a much higher degree of overlap at the LD block resolution. As in the LDL analysis, we visualized the model fitting results at variant level in Figure~\ref{fig:manhatten_BMI}, and provided more details about the top five variants with largest absolute effect size in mPRS and vPRS in Table \ref{tab:top_variants}. All top five variants in mPRS and the top three variants in vPRS have been identified in the literature to be associated with BMI  \cite{huang2022genomics,wood2016variants,sidorenko2024genetic,zhou2023interaction,felix2016genome,pulit2019meta}, and are mapped to genes well-known to be associated with BMI or obesity such as \emph{FTO}, \emph{MC4R} and \emph{DCC}. The other two top variants in vPRS are novel findings and have not been reported in existing literature as relevant for BMI. Their underlying biological pathways need further investigation.

\begin{table}
\caption{Top five variants in mPRS and vPRS selected by snpboostlss for BMI. Their rsID, mapped genes (GRCh37/hg37) and association to BMI in existing literature are reported.}
\label{tab:top_variants}
\begin{tabular}{cccc|cccc}
\toprule
\multicolumn{4}{c|}{mPRS}   & \multicolumn{4}{c}{vPRS}     \\ \hline
Rank & SNP        & Gene              & BMI-related & Rank & SNP         & Gene              & BMI-related \\ \hline
1    & rs62106258 & \emph{LINC01865, LINC01874}      & Yes\cite{huang2022genomics}         & 1    & rs62106258  & \emph{LINC01865, LINC01874}      & Yes\cite{huang2022genomics}         \\
2    & rs1421085  & \emph{FTO} & Yes\cite{wood2016variants}         & 2    & rs1421085   & \emph{FTO} & Yes\cite{wood2016variants}         \\
3    & rs2229616  & \emph{MC4R}              & Yes\cite{sidorenko2024genetic}         & 3    & rs116873887 & \emph{LINC01630,
DCC}      & Yes\cite{zhou2023interaction}  \\
4    & rs543874  & \emph{LINC01741, SEC16B}              & Yes\cite{felix2016genome}         & 4    & rs1017618 & \emph{LINC01923, SATB2}      & No \\
5    & rs13107325  & \emph{SLC39A8}           & Yes\cite{pulit2019meta}         & 5    & rs117895800 & \emph{LINC00424, LINC00540}      & No          \\ \bottomrule
\end{tabular}
\end{table}

\begin{figure}
    \centering
    \includegraphics[width=\linewidth]{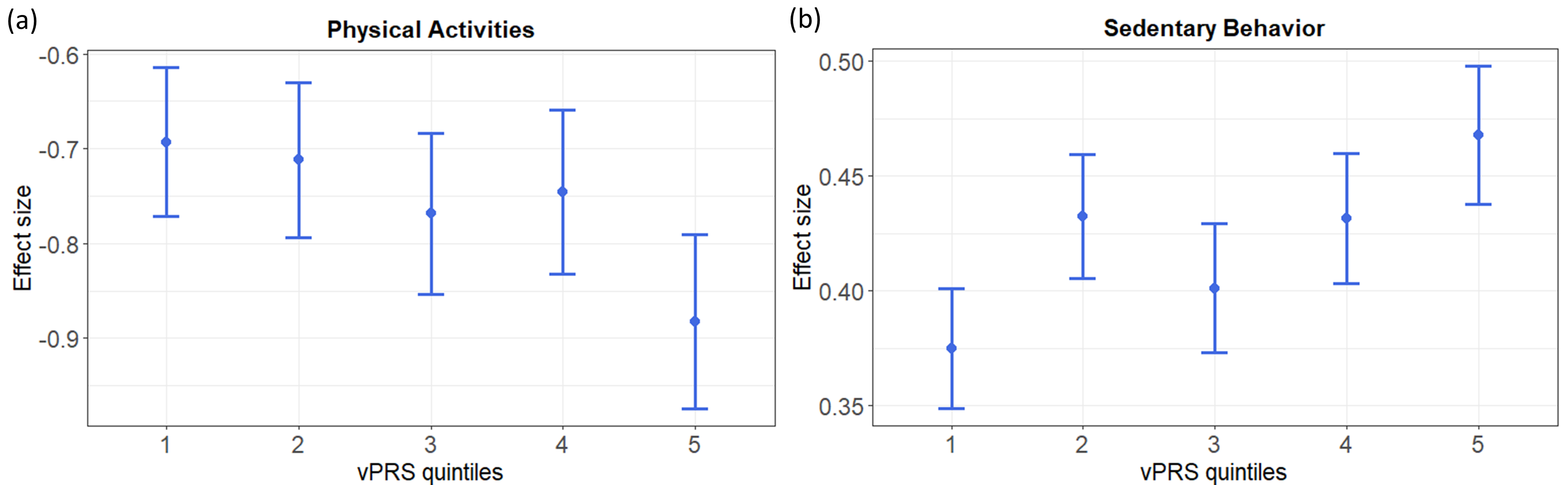}
    \caption{Interaction effects between vPRS and environmental factors on BMI in UK Biobank. (a) Effect size of physical activity on BMI by vPRS quintiles; (b) Effect size of sedentary behavior on BMI by vPRS quintiles. For each quintile, the estimated effect size along with 95\% CI is displayed.}
    \label{fig:BMI_quintile}
\end{figure}

We then investigated whether the constructed vPRS can show interaction effects with environmental factors in the test set. The environmental factors considered here are physical activity (PA) and sedentary behavior (SB) (\emph{Methods}). The main effect of physical activity is significantly negative ($P < 2\times 10^{-16}$, \emph{SI} Table S2), which is consistent with the expectation that more activity in general leads to lower BMI. In addition, we observed a significantly negative vPRS-PA interaction ($P=8.73\times 10^{-4}$ , \emph{SI} Table S2). That means subjects with higher vPRS have more negative total effects of physical activity (Figure~\ref{fig:BMI_quintile}(a)), so they could benefit more, in terms of lowering BMI, from doing e.g., additional sports. When sedentary behavior is considered as the environmental factor, we found both its main effect ($P < 2\times 10^{-16}$, \emph{SI} Table S2) and vPRS-SB interaction effect ($P=1.31\times 10^{-3}$ , \emph{SI} Table S2) to be significantly positive, meaning that people with longer sitting time have higher BMI on average, which is again as expected. Also subjects with higher vPRS have larger total positive effect of sedentary behavior (Figure~\ref{fig:BMI_quintile}(b)), so reducing their sitting time can be more beneficial in terms of lowering BMI. Both interactions remained significant after we adjusted for additional vPRS-covariate interaction terms in the model ($P=1.33 \times 10^{-3}$  and $2.72 \times 10^{-5}$ for PA and SB, respectively, \emph{SI} Table S2). In summary, our constructed vPRS shows once again significant interaction effects with relevant environmental factors. This demonstrates the potential use of the vPRS constructed by snpboostlss to stratify individuals based on their genetic liability towards benefits from lifestyle changes.

\section*{Discussion}

Polygenic risk scores provide an estimate for the genetic predisposition of each individual on the phenotype of interest. Most existing methods focus on predicting the mean of the trait, and only some vQTL methods estimate genetic effects on the phenotypic variability \cite{conley2018sibling, johnson2022polygenic, miao2022quantile}. But mean and variance are handled separately by these methods. In this work, we introduced snpboostlss, which implements the batch-wise cyclical gradient boosting for Gaussian location-scale models on large scale genetic data. With the proposed snpboostlss method, we are for the first time able to develop mPRS and vPRS simultaneously and also intrinsically capture the mutual influence between the two scores. Our simulation studies demonstrate that, in the case of genetically induced heteroscedasticity, snpboostlss can yield accurate variant selection and prediction for mPRS, which could be beneficial for downstream understanding of biological mechanism as well as patient risk stratification. Also our vPRS derived using genotype and baseline phenotype data can provide estimates of the within-individual variability comparable to the benchmark longitudinal estimator. Therefore, we would like to advocate that the derived vPRS from snpboostlss can be considered as an efficient proxy for individual phenotypic variability, especially when longitudinal observations are limited.

Moreover, our method advances the identification of GxE interactions for complex traits. Evidence suggests that genetics, environments, and their ubiquitous interactions jointly shape human phenotypes\cite{manolio2009finding}. But identifying variants involved in GxE interactions in complex trait research still remains a challenging task. The applications of our method on UK Biobank data with LDL and BMI as the phenotypes of interest demonstrate that our constructed vPRS leads to significant interaction effects with various relevant environmental factors like use of statin medication  for LDL or physical activity and sedentary behavior for BMI. These results illustrate the potential use of the snpboostlss as an effective tool to identify variants that are potentially involved in GxE interactions. In addition, the constructed vPRS could be used in practice to stratify individual sensitivity towards environmental changes, so that clinicians could understand much clearer which patient cohorts could benefit more from medical intervention or lifestyle changes. 

Despite the presented promising results, the proposed method also has some limitations. First, our approach inherits some limitations from statistical boosting. Boosting does not provide closed formulas for standard errors of coefficient estimates, i.e., statistical inference is not directly possible. Second, we demonstrated that vPRS constructed by snpboostlss shows significant GxE interactions, but being involved in GxE interactions is only sufficient but not necessary for a variant to be included in vPRS. Therefore, the vPRS may also capture other mechanisms that can lead to heteroscedasticity, such as gene–gene interactions and genetic effects on higher moments of the phenotypic distribution. Therefore, results based on vPRS need to be closely investigated further and interpreted with caution. Thirdly, when verifying GxE interactions for LDL in parallel group design, we mimicked a randomized controlled trial with observational data (target trial emulation). However, there is always an unavoidable gap between an actual RCT and the observational data even after adjusting for confounders. For example, we must rely on the assumption of no omitted confounders in the observational study. Our analysis also assumes that at the first revisit, subjects taking statins are already on stable use of the medication, thus the LDL measurements in the database can properly reflect the effects of the medication. Fourthly, via the UK Biobank application we have shown that the vPRS derived by snpboostlss is a good proxy for the genetic component in GxE interactions. However, we also observed that if we use mPRS as the genetic component in our examples, the significance of GxE interactions often remains, which was also observed by previous research\cite{natarajan2017polygenic}. Further study is needed to understand the different roles mPRS and vPRS play in GxE interactions and to explore how to improve patient stratification through potentially joint use of both scores.     

In future research, we will further explore other potential use of our method in GxE studies. For example, our construction of vPRS is environmental factor free, i.e., if we consider vPRS as a reasonable representation of the G component in GxE interactions, we could use it to test which environmental factors have significant interaction effects with genes. In addition, constructing proper measurements for certain environmental factors could be challenging. Our vPRS may also help to validate the measurements of environmental factors known to interact with genetic factors.

In addition, our method constructs mPRS and vPRS simultaneously. In the future, we plan to take advantage of these new insights from the location-scale models to improve and extend PRS predictions in general. One potential direction is to go beyond the classical point-prediction of PRS towards genotype-based individual prediction intervals for continuous phenotypes. The main advantage of prediction intervals is that they can report the involved statistical uncertainty and might help clinicians also in the communication of risks with patients.

Furthermore, we could take advantage of the modular structure of boosting to model more complex biological phenomena. We will incorporate different loss functions to extend the snpboostlss framework to be applicable also to other kinds of phenotypes such as recurrent event count data and failure time in the framework of distributional regression. Apart from enabling new loss functions in the framework, we could also alter the base-learners. For example, non-linear base-learners could be adopted to capture dominant or recessive hereditary schemes.

To conclude, this paper introduces distributional regression for the first tome to the field of polygenic risk scores. It successfully achieves simultaneous and efficient construction of mPRS and vPRS, and demonstrates the application of the vPRS in gene-environment interaction studies. It hints at the clinical use of vPRS in personalized intervention, namely to determine intervention measures based on individual characteristics of patients including their genetic liability towards changes in lifestyle, medication or other environmental factors.

\section*{Methods}
\label{sec:methods}

\subsection*{Statistical methods}

For each individual $i=1,...,n$ we observe the phenotype outcome $y_i$ and  $p$ genetic variants $g_{i,j}$ for $j=1,...,p$. The genetic data of $n$ individuals are given in the genotype matrix $\bm{G}=(g_{i,j})\in[0,2]^{n\times p}$. Considering a Gaussian location-scale model on a continuous phenotype, we use the following notation 
\begin{equation}\label{eqn:gaussian_ls}
   \hspace{2in} y_i \overset{\mathrm{ind.}{}}{\sim} N(\mu_i, \sigma_i^2),\quad\mu_i=\bm{x}_i'\bm{\beta}, \quad \log(\sigma_i)=\bm{z}_i'\bm{\gamma}, 
\end{equation}
where $\bm{x}_i$ and $\bm{z}_i$ are subsets of $\bm{g}_i = (g_{i,1}, \dots, g_{i,p})'\in[0,2]^{p}$ which corresponds to the genotype data of individual $i$. Our methodological aim is to identify $\bm{x}_i$ and $\bm{z}_i$ and estimate their corresponding coefficients $\hat{\bm{\beta}}$ and $\hat{\bm{\gamma}}$ via minimizing the loss function defined as the negative log-likelihood  -- which is equivalent to maximizing the likelihood. 

An effective tool to perform variable selection and coefficient estimation simultaneously for statistical models in the presence of potentially high-dimensional data is component-wise gradient boosting\cite{buhlmann2007boosting,mayr2018boosting}. Gradient boosting requires the specification of a loss function $\rho(\bm{y}, \bm{\hat{y}})$ and the so-called base-learners. In order to estimate statistical models with additive structure, separate regression-type base-learners $h_j, j=1,\dots , p$ can be used for each single variable (\emph{statistical boosting}\cite{mayr2014evolution, mayr2014extending}) that are iteratively fitted to the negative gradient of the loss function. Starting at iteration $m=0$ with a starting value $\bm{\hat{y}}^{(0)}$, the following steps are repeated until a maximum number $m_{\mathrm{stop}}$ of boosting iterations is reached\cite{buhlmann2007boosting}:
\begin{enumerate}
    \item Set $m:=m+1$ and compute the negative gradient of the loss function: 
    $$\bm{u}^{(m)}=\left.-\dfrac{\partial\rho(\bm{y},\bm{\hat{y}})}{\partial\bm{\hat{y}}}\right\vert_{\bm{\hat{y}}=\hat{\bm{y}}^{(m-1)}}$$
    \item Fit every base-learner $h_j$ separately to the negative gradient $\bm{u}^{(m)}$ and select the best fitting base-learner $\hat{h}^{(m)}_{j^*}$,
    \item Update the predictor with a learning rate $\nu\geq0$: $\hat{\bm{y}}^{(m)}=\hat{\bm{y}}^{(m-1)}+\nu\hat{h}^{(m)}_{j^*}$
    \item Stop if $m=m_{\mathrm{stop}}$
\end{enumerate}
To fit a generalized additive model for location, scale and shape (GAMLSS)\cite{rigby2005generalized} which includes Gaussian location-scale model as a special case, a cyclical update approach on different distribution parameters\cite{mayr2012generalized} can be further adopted. 

When working on genetic data from large cohort studies we face not only a high-dimensional setting with $p>n$ but also a large-scale setting with large $n$ and large $p$. Large-scale settings often lead to extended computation times as well as memory issues. To overcome these challenges and apply statistical boosting directly on individual genotype data, Klinkhammer et al.\cite{klinkhammer2023statistical, klinkhammer2024genetic} developed the snpboost algorithm for mean regression models which incorporates an additional batch-building procedure before the boosting iterations. Consequently, boosting is performed only on a small subset of variants, thus largely improving computational efficiency. 

We extended this framework to Gaussian location-scale models by introducing a batch-building procedure in the cyclical boosting approach for GAMLSS. Our proposed snpboostlss algorithm is able to perform variant selection and effect estimation for both mean and variance parameters simultaneously, while maintaining computational efficiency for large genetic data.

The new snpboostlss algorithm is summarized in Figure~\ref{fig:algorithm_flowchart} and its details are given in Section S1 of the supplementary information. The algorithm consists of two parts, an outer loop (shown in blue in Figure~\ref{fig:algorithm_flowchart}) and an inner loop (shown in grey in Figure~\ref{fig:algorithm_flowchart}). The outer loop corresponds to the batch-building procedure, where we extract the $p_{\mathrm{batch}}$ variants ($p_{\mathrm{batch}}\ll p$) with highest correlation to the current negative gradient of the loss function with respect to $\mu$ and $\sigma$ to form separate batches for $\mu$ and $\sigma$, respectively. Then we enter the inner loop to sequentially update coefficients for $\mu$ and $\sigma$ via cyclical boosting on those constructed variant batches for a maximum number of $m_{\mathrm{batch}}$ iterations. Early stopping of boosting within a given batch (i.e., not completing all $m_{\mathrm{batch}}$ iterations) for either $\mu$ or $\sigma$ is allowed if there exists a variant outside the batch showing higher correlation with the negative gradient vectors than all variants inside the batch: In this case a variant outside the batch may provide a better fit to the current negative gradient vector. If boosting is stopped early for either $\mu$ or $\sigma$, the other parameter will keep being updated until the stopping criteria for the inner loop has been met. The inner loop is terminated when either both parameters are early stopped or the maximum number of boosting iterations is reached. Once the inner loop has been completed, we return to the outer loop to rebuild batches and repeat the process. In total, we fit a maximum of $b_{\mathrm{max}}$ batches or stop the algorithm early if the fitted model cannot show performance improvements on a validation set for $b_{\mathrm{stop}}$ consecutive batches. The stopping iteration is chosen as the one in which the loss evaluated on validation set reaches its minimum, which in our case is equivalent to the maximum of the predictive likelihood.

\begin{figure}[ht]
\centering
\includegraphics[width=\linewidth]{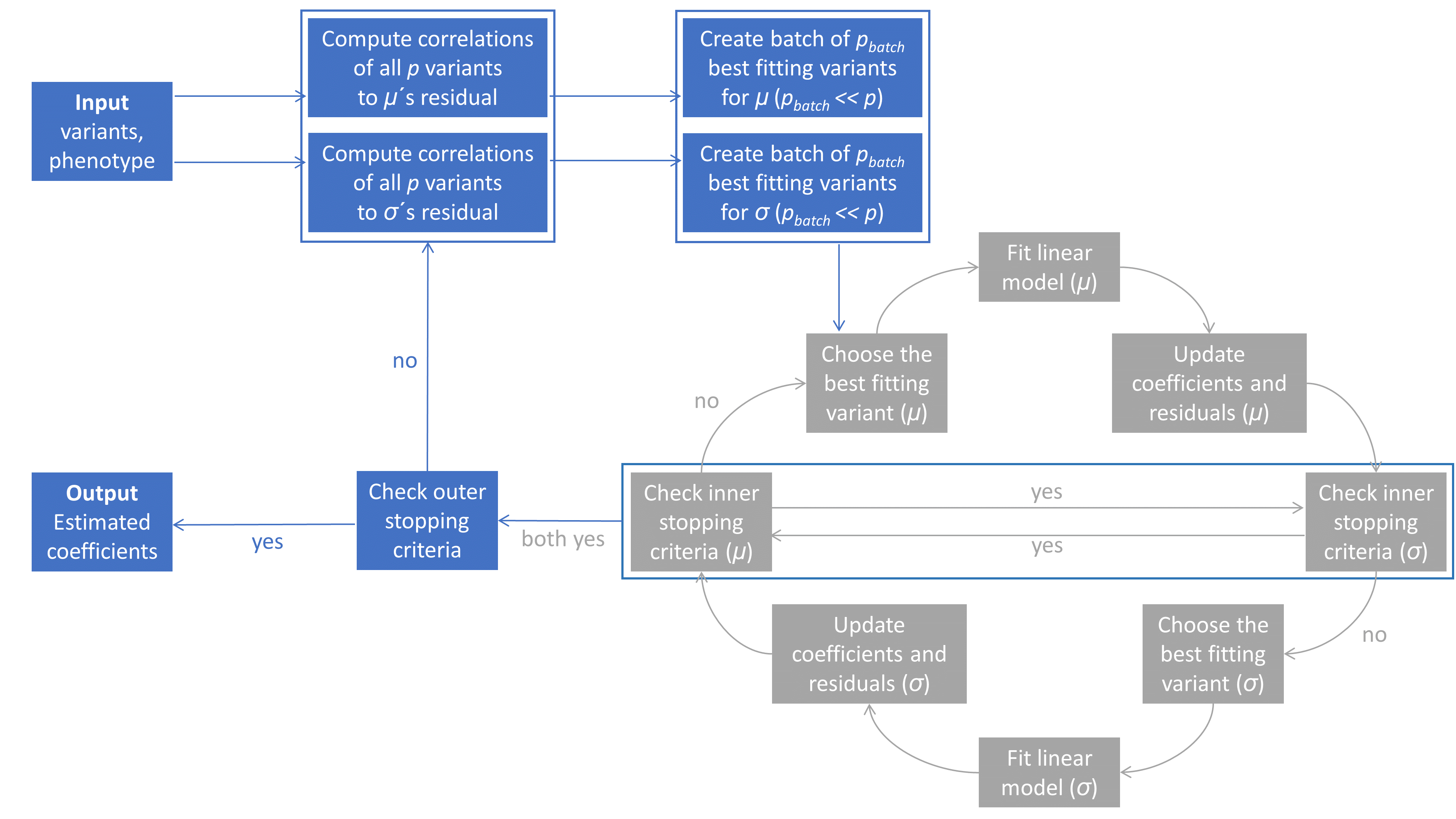}
\caption{Workflow of the new snpboostlss algorithm. It consists of an outer loop (in blue) related to variant batch creation and overall stopping criteria evaluation  and an inner loop (in grey) representing the model fitting via boosting on given batches.}
\label{fig:algorithm_flowchart}
\end{figure}

As shown in the Step~3 of the general description of the boosting algorithm, the learning rate $\nu$ determines the step length moving from starting value towards optimum in the boosting algorithm and is usually predefined at a fixed value. However, as reported in Zhang et al. (2022)\cite{zhang2022adaptive}, for complex models with several distributional parameters such as the Gaussian location-scale model, different distributional parameters may refer to different scales regarding their impact on the gradient. Using a fixed learning rate in such cases might lead to imbalanced updates of parameters, which prevents some sub-models to be sufficiently fitted within a limited number of boosting iterations. To overcome this issue, we followed the recommendation from Zhang et al.\cite{zhang2022adaptive} and added the option of an adaptive step length in our algorithm. This allows the learning rate to be adapted in different iterations according to the parameter scale. Details on the calculation of adaptive step lengths and our simulation results on its effect can be found in Section S2.1 in the supplementary information. The implementation is provided on GitHub (https://github.com/boost-PRS/snpboostlss).

\subsection*{Simulation settings}

We conducted simulation studies to investigate the behavior of the proposed snpboostlss algorithm in various controlled data generating scenarios. The simulation studies aim at two main goals: first, to compare the performance of snpboostlss with snpboost on estimating mPRS; second, to compare the derived vPRS with within-individual variability estimator using longitudinal data.

Simulations were based on HAPNEST synthetic genotype data \cite{wharrie_hapnest_2023} combined with simulated phenotypes. We focused on variants from Chromosome 22, which contains in total $106,904$ SNPs in the HAPNEST data. In each simulation, we randomly selected $p=20,000$ variants with a similar correlation structure (linkage disequilibrium) compared to the original genotype data. $n=20,000$ individuals were randomly selected and split into 50\% training, 20\% validation and 30\% test sets.

Continuous phenotypes were simulated based on the Gaussian location-scale model in Equation~\eqref{eqn:gaussian_ls}. The number of informative variants depends on a predefined sparsity level $s_{\mu}=s_{\sigma}:=s\in \{0.1\%, 1\%\}$. In each simulation, $s$ of the 20,000 variants were first randomly selected to be informative for $\log(\sigma)$, whose coefficients were generated from $U(-0.25, 0.25)$. Using such a small range is to ensure the magnitude of the scale parameter to fall into a reasonable range. Afterwards another $s$ of the 20,000 variants were randomly selected to generate $\mu$ with their effect sizes sampled from
$$N\left(0, \frac{\frac{\bar\sigma^2}{1-h^2}h^2}{s\cdot p}\right), \text{ where } \bar\sigma^2=\frac{\sum_i\sigma_i^2}{n}, \text{ and } h^2\in\{0.1, 0.7\}.$$
This way of data generation is similar to that in Priv{\'e} et al. (2019) \cite{prive2019efficient} and provides datasets with average heritability achieving our desired level $h^2$ being 0.1 or 0.7. As a result, our simulation study is able to account for different genetic architectures by considering different combinations of heritability $h^2$ and sparsity $s$. With the above generated coefficients, we obtained the true values of $\mu_i$ and $\sigma_i$. We then randomly sampled 100 phenotype values for the $i$-th individual from $N(\mu_i, \sigma_i)$ independently. We randomly chose one as the baseline measurement and the other 99 as repeated measurements. Altogether they form our longitudinal data. Under each genetic architecture, we simulated 100 different datasets. 

PRS models for the mean were derived by snpboost (with default settings; for more details see Klinkhammer et al.  \cite{klinkhammer2023statistical}) and for both mean and variance by snpboostlss (with same parameter setting as snpboost and adaptive step length in addition) using the baseline data of the training and validation sets. We compared the effect of using adaptive step length versus traditional fixed step length of value 0.1 in snpboostlss, and found that adaptive step length can achieve better prediction performance, more balanced updates between parameters and higher variable selection accuracy (detailed results are provided in Section S2.1 in the supplementary information). Therefore, we set adaptive step length as the default choice for our snpboostlss implementation and also used it for the rest of this paper. 

The performance of PRS models were evaluated on the test set by various metrics regarding their predictive performance, accuracy of variant selection and computation time. In detail, the predictive performance was measured by the $R^2$ for the mean defined as squared correlation between the predicted and true phenotype values \cite{staerk2024generalizability}, or predictive loss defined as negative log-likelihood on test data which takes both $\mu$ and $\sigma$ into account. Regarding variant selection accuracy, we calculated the percentage of included variants in the final model, true positive rate, and true negative rate. We also performed sanity checks on the performance of snpboostlss. Results can be found in Section S2.2 in the supplementary information. Besides estimation of within-individual variability via vPRS, another estimator for $\sigma_i$ is given by the sample standard deviation of the longitudinal observations of the $i$-th subject. The estimation accuracy of vPRS was further compared with that of the longitudinal data based estimator via the correlation between predicted and true values of $\sigma$ on the test set. Simulations were run on a high performance computing cluster at Marburg University. For each simulation, 2 CPUs with 12 GB memory per CPU were used. The code to reproduce the results can be found on GitHub (https://github.com/boost-PRS/snpboostlss).  

\subsection*{UK Biobank data processing and analysis}

We analyzed data from the UK Biobank (UKBB) database under Application Number 135122. The UK Biobank is a large-scale prospective cohort study including more than half a million participants from the United Kingdom aged between 40 and 69 years old when recruited \cite{bycroft2018uk}. The database comprises genome-wide genotype data at individual level and various in-depth phenotypic information such as biological measurements, medication status as well as lifestyle information. 

We chose low-density lipoprotein (LDL, UKBB field 30780) and body mass index (BMI, UKBB field 21001) as our phenotypes of interest, because they are typical examples of phenotypes being influenced by both genetic and environmental factors. Our objectives are to implement snpboostlss to construct mPRS and vPRS for LDL and BMI respectively, to compare the variants included in mPRS and vPRS for each phenotype and to investigate potential GxE interactions. 

For each trait, we removed participants with conflicting genetic sex (UKBB field 22001) and self-reported sex (UKBB field 31), filtered for unrelated individuals (UKBB resource 668) with self-reported white British ancestry (UKBB field 21000) and availability of baseline phenotype data, resulting in $n = 244,583$ and $n = 351,891$ subjects for LDL and BMI, respectively. We randomly divided the data into training, validation and test sets with allocation 2:1:1. We used genome-wide genotype data and filtered for variants with a genotyping rate of at least 90\% and a minor allele frequency of at least 0.1\%, resulting in $p = 604,967$ and $p = 510,061$ biallelic genetic variants on autosomes for LDL and BMI, respectively. 

We applied snpboostlss on the training and validation sets with default parameter settings. The selected variants were assigned to approximately independent LD-Blocks, defined as 1,703 genomic regions of high linkage disequilibrium in the European population \cite{berisa2015approximately}. To achieve this, the genomic co-ordinates of the selected variants were intersected with the co-ordinates of the predefined set of LD-Blocks. The top 5 variants with the largest absolute effect sizes in mPRS and vPRS models are mapped to genes based on Genome Reference Consortium Human Build 37 (GRCh37/hg37) and checked for their association with the interested trait in GWAS Catalog\cite{Cerezo_2024}.

\subsection*{Detection of GxE interactions using baseline data}

As discussed in \emph{Introduction}, the vPRS, an aggregated summary of variants affecting phenotypic variability, gives potential genetic information in GxE interactions. We aimed to test whether the vPRS constructed by snpboostlss can show an interaction effect with relevant environmental factors. For LDL, the environmental factor was the usage status of any statins (UKBB field 20003), which is one class of common prescription drugs used to lower LDL. For BMI, the environmental factors we considered were physical activity (PA, based on UKBB fields 864, 874, 884, 894, 904 and 914) and sedentary behavior (SB, based on UKBB fields 1070, 1080 and 1090). Details about the construction of PA and SB can be found in existing literature\cite{marderstein2021leveraging,wang2019genotype}. For PA, we assigned a three-level categorical score (low, medium, and high) according to the International Physical Activity Questionnaire Guideline. We defined SB as the total time (hours) per week spent on driving, using a computer, and watching television. 

To test vPRS$\times$E interaction effects, we fitted the following linear model on the test set:
$$\text{Y}_i \sim \text{mPRS}_i + \text{vPRS}_i + \text{E}_i + \text{vPRS}_i\times \text{E}_i$$
where $\text{Y}_i$ is the phenotype of interest, $\text{mPRS}_i$ is the mPRS developed by snpboostlss, $\text{vPRS}_i$ is the standardized vPRS from snpboostlss with mean 0 and variance 1, and $E_i$ is the environmental factor for the $i$-th individual. We further adjusted for age (UKBB field 21022), sex (UKBB field 31), genotyping array (UKBB field 22000), and top 12 PCs (UKBB field 22009). To check the robustness of our results, we repeated our vPRS$\times$E analysis by fitting the model above with vPRS-age and vPRS-sex as additional covariates\cite{keller2014gene}. To verify the potential interaction effects, we further divided the test set into 5 quintiles based on the vPRS and compared estimated effect of the environmental factor across vPRS quintiles.

\subsection*{Verification of GxE interactions with a self-controlled design}

In the LDL application, we further verified the GxE interaction using repeated observations on LDL and statins usage status with a self-controlled design. The repeated observations are those from the initial visit (serving as baseline) and first revisit in UKBB. We measured the effect of statins by the changes in LDL from baseline measurement to the first revisit. We focused on the people in the test set who did not take statins at baseline but were taking statins at first revisit and had LDL measured at both visits. This filtering process leads to a sample of 767 subjects (\emph{SI}, Figure S4). We then investigated whether people in high-vPRS group experienced larger LDL decrease than low-vPRS group. High/low vPRS groups were defined as subjects with vPRS beyond 75\%/25\%  or 90\%/10\% quantile of vPRS in the complete test set. Two-sample t-test was performed to compare the change from baseline in LDL between high-vPRS and low-vPRS groups.

\subsection*{Verification of GxE with a parallel-group design}

In the LDL application, we also verified the GxE interaction using repeated observations on LDL and statins usage status mimicking a parallel-group design. We filtered the test set for subjects who had baseline LDL higher than 3.36 mmol/L (130 mg/dl)\cite{adams2014rosuvastatin, zhao2018efficacy, talavera2013double, florentin2013colesevelam, her2010effects}, were not taking statins at baseline and had repeated measurements on LDL and statins status at both baseline and first revisit. This filtering process leads to a verification set with 1,276 eligible subjects, among which 530 belong to the intervention group (taking statins at first revisit) and 746 belong to the control group (not taking statins at the first revisit). Details of filtering process can be found in Figure S4 in supplementary information.

One thing worth noting is the LDL threshold we used to identify subjects eligible for our analysis. The threshold is crucial because it influences the sample size. The threshold 3.36 mmol/L is a commonly used eligibility criteria in trials with statins as primary prevention of cardiovascular disease \cite{adams2014rosuvastatin, zhao2018efficacy, talavera2013double, florentin2013colesevelam, her2010effects}. We are aware that there are other thresholds used in previous statins trials, such as 1.81 mmol/L, 2.58 mmol/L and 4.14 mmol/L\cite{adams2014rosuvastatin}. We did not choose the lower thresholds since they are often adopted in trials where patients already experienced severe or acute cardiovascular disease in the first place and statins were used as secondary preventative measures \cite{ballantyne2013alteration, pitt2012comparison, pitt2008lipid, bellia2010early}. We did not adopt the higher threshold (4.14 mmol/L) because it is often used as the threshold for general population to be considered as high LDL \cite{ballantyne2003efficacy, brown200252, celik2012effects}. But our test set has an average age of 57, which is relatively old and may increase the risk of cardiovascular diseases and the prevalence of other chronic diseases. Therefore we believe that a moderately high threshold (3.36 mmol/L) is more appropriate as the eligible criteria for our analysis. For completeness, we also performed the same analysis with other thresholds. See \emph{SI}, Figure S6 and S7 for more results.  

We implemented inverse probability of treatment weighting \cite{chesnaye2022introduction} to adjust for potential confounders and to mitigate the selection bias in the observational data. We identified potential confounders based on previous statins trials\cite{ballantyne2013alteration, pitt2008lipid, pitt2012comparison, betteridge2007effects} and the national guidance for lipid management in UK \cite{cegla2023national}. The confounders we adjusted for are the baseline values of age (UKBB field 21022), sex (UKBB field 31), BMI (UKBB field 21001), low-density lipoprotein (UKBB field 30780), high-density lipoprotein (UKBB field 30760), C-reactive protein (UKBB field 30710), triglycerides (UKBB field 30870), apolipoprotein B (UKBB field 30640), smoking (UKBB fields 1239, 20116), diabetes (UKBB field 2443) and systolic blood pressure (UKBB field 4080). We fitted a logistic regression model to calculate the probability of being exposed to intervention (i.e., propensity score) given an individual's characteristics of the above confounders. Then weight is calculated for each individual as 1/(propensity score) for those in the intervention group and 1/(1-propensity score) for those in the control group. Incorporation of these weights aims at creating a pseudo-population in which confounders are equally distributed across two treatment groups. 

Our endpoint is the change from baseline in LDL. The estimated treatment effect of statins therapy is then given by the difference between intervention and control groups regarding change in LDL from baseline. We first descriptively illustrated the difference in treatment effect of statins therapy between high and low vPRS groups in Figure~\ref{fig:LDL_application}(c). Each point in the plot represents the weighted average of change from baseline in LDL for the corresponding vPRS-treatment-subgroup where the weights are obtained from the IPTW calculation. As such, the slope of each line represents the treatment effect in the corresponding vPRS-subgroup, and the difference between the slopes of two lines represents the interaction effect between vPRS and statins. Complete results based on different eligibility criteria and different high/low vPRS subgroups can be found in \emph{SI} Figure S6. 

We further quantified the overall treatment effect by fitting the following linear regression model:
\begin{equation}\label{eqn:statin_adj_effect}
    \Delta\text{LDL}_i \sim \text{statins.1}_i + \text{LDL.0}_i + \text{age}_i + \text{sex}_i + \text{PC1}_i + \cdots + \text{PC12}_i
\end{equation}
where $\Delta\text{LDL}_i=\text{LDL.1}_i-\text{LDL.0}_i$ represents the change in LDL from baseline (LDL.0) to first revisit (LDL.1), statins.1 is the binary variable describing whether a subject was taking statins at first revisit, so it represents the treatment group and its coefficient quantifies the treatment effect of statins. In addition we adjusted for baseline LDL, age, sex and top 12 principal components. This model was fitted via weighted linear regression with weights derived from IPTW. We performed the same analysis in all vPRS-based subgroups and visualized the treatment effects in a forest plot (Figure~\ref{fig:LDL_application}(d)). To investigate whether there is significantly different treatment effects in vPRS-based subgroups, we implemented the subgroup interaction analysis. Specifically, we added an interaction term between the binary vPRS grouping variable (high/low) and $\text{statins.1}$ to Model \eqref{eqn:statin_adj_effect} and focused on whether the interaction term is significant or not. More comprehensive results based on different eligibility criteria and different high/low vPRS subgrouping can be found in \emph{SI} Figure S7.

\section*{Data availability}

The data analyzed in this study is subject to the following licenses/ restrictions: This research has been conducted using the UK Biobank resource under application number 135122 (http://www.ukbiobank.ac.uk). Requests to access these datasets should be directed to UK Biobank, http://www.ukbiobank.ac.uk.

\section*{Code availability}

An R implementation of snpboostlss and the code for simulation studies and real data applications are provided in GitHub (https://github.com/boost-PRS/snpboostlss). 

\bibliography{main}



\section*{Acknowledgements}

The work on this article was supported by the Deutsche Forschungsgemeinschaft (DFG, grant number 534238115).

\section*{Author contributions statement}

QW, HK, AM, CM, and CS contributed to conception and design of the method. QW wrote the code, performed the experiments and wrote the first draft of the manuscript. KK performed the gene mapping and enrichment analysis. All authors contributed to manuscript revision, read and approved the submitted version.

\section*{Competing interests}

The authors declare no competing interests.




\end{document}


\maketitle

\newpage

\section{\emph{snpboostlss} algorithm}

\begin{algorithm}[!ht]
		\caption{\textbf{SNPBOOSTLSS}}\label{alg:snpboostlss}
		\KwData{%
			\begin{tabular}[t]{@{}l@{\hspace{2em}}l@{}}
				Phenotype data:            & \( \bm{y} \in \mathbb{R}^n, \) \\
				Genotype data:             & \( \bm{G}=(g_{i,j}) \in [0,2]^{n \times p}, \) \\
                Learning rate:               & \( \nu \geq 0, \) \\
				Batch size:                  & \( p_{\text{batch}} \in \{1,\cdots, p\}, \) \\
				Max. number of boosting iterations per batch: & \( m_{\text{batch}} \in \mathbb{N}, \) \\
				Max. number of batches:             & \( b_{\text{max}} \in \mathbb{N}, \) \\
				Stopping lag for outer stopping criterion:                & \( b_{\text{stop}} \in \mathbb{N}. \)
			\end{tabular}
		}
		\KwResult{
			\begin{enumerate}
				\item \textbf{Initialization:} \\
                Set boosting index $m=0$. \\
                Initialize $\hat{\boldsymbol{\beta}}^{(0)}=(\bar{y}, 0, \cdots, 0)'$, $\hat{\boldsymbol{\gamma}}^{(0)}=(\log(s_y), 0, \cdots, 0)'$ where $s_y$ is the sample standard deviation of $\bm{y}$. \\
                Calculate residuals: \\
                $\boldsymbol{r}_{\mu}^{(0)}=\left [ \frac{y_i-\boldsymbol{g}_i'\hat{\boldsymbol{\beta}}^{(0)}}{\exp(2\boldsymbol{g}_i'\hat{\boldsymbol{\gamma}}^{(0)})} \right ]_{i=1,\cdots,n}$ and $\boldsymbol{r}_{\sigma}^{(0)}=\left [ \frac{(y_i-\boldsymbol{g}_i'\hat{\boldsymbol{\beta}}^{(0)})^2}{\exp(2\boldsymbol{g}_i'\hat{\boldsymbol{\gamma}}^{(0)})}-1 \right ]_{i=1,\cdots,n}$
            
				\item \textbf{Outer loop:}
                Set outer counter $k=1$
				\begin{enumerate}
					\item[(a)] \textbf{Screening}:\\
				    \begin{enumerate}
				        \item[(1)] \textbf{Batch building for $\mu$}:\\
                        Compute correlations $c_{\mu j}^{(m)}=\rho(\boldsymbol{r}_{\mu}^{(m)}, \boldsymbol{g}_j)$, $j=1,\cdots, p$.\\
                        Create batch $B_{\mu k}$ of $p_{\text{batch}}$ variants with highest absolute correlations $|c_{\mu j}^{(m)}|$. \\
                        Save the highest absolute correlation outside the batch as $c_{\text{stop},\mu}=\max_{j\notin B_{\mu k}}|c_{\mu j}^{(m)}|$. \\
                        Set early stopping flag $F_{\text{stop},\mu}=\texttt{FALSE}$.
                        \item[(2)] \textbf{Batch building for $\sigma$}:\\
                        Compute correlations $c_{\sigma j}^{(m)}=\rho(\boldsymbol{r}_{\sigma}^{(m)}, \boldsymbol{g}_j)$, $j=1,\cdots, p$.\\
                        Create batch $B_{\sigma k}$ of $p_{\text{batch}}$ variants with highest absolute correlations $|c_{\sigma j}^{(m)}|$. \\
                        Save the highest absolute correlation outside the batch as $c_{\text{stop},\sigma}=\max_{j\notin B_{\sigma k}}|c_{\sigma j}^{(m)}|$. \\
                        Set early stopping flag $F_{\text{stop},\sigma}=\texttt{FALSE}$.
				    \end{enumerate}
					\item[(b)] \textbf{Inner loop}: Set inner counter $l=1$
					\begin{enumerate}
						\item[(1)] If $l>m_{\text{batch}}$, end inner loop and go to (c); else proceed to (b)(2).
						\item[(2)] 
                        Calculate inner loop stopping flag $F_{\text{stop,inner}}=F_{\text{stop},\mu}\times F_{\text{stop},\sigma}$. \\
						If $F_{\text{stop,inner}}=\texttt{TRUE}$, end inner loop and go to (c); \\
                        else $m:=m+1$ and proceed to (b)(3). 
					\end{enumerate}
				\end{enumerate}

		\end{enumerate}
		}
	
\end{algorithm}

\newpage

\begin{algorithm}[!ht]
\caption*{\textbf{SNPBOOSTLSS (Continued)}}
\addcontentsline{loa}{algorithm}{My Algorithm}
        \KwResult{
            \begin{enumerate}
                \item[] 
                \begin{enumerate}
                    \begin{enumerate}                        
                        \item[(3)] For $\mu$: \\
                        \begin{enumerate}
                            \item [(i)] If $F_{\text{stop},\mu}=\texttt{TRUE}$, \\
                            $\hat{\boldsymbol{\beta}}^{(m)}=\hat{\boldsymbol{\beta}}^{(m-1)}$, $\boldsymbol{r}_{\mu}^{(m)}=\left [ \frac{y_i-\boldsymbol{g}_i'\hat{\boldsymbol{\beta}}^{(m)}}{\exp(2\boldsymbol{g}_i'\hat{\boldsymbol{\gamma}}^{(m-1)})} \right ]_{i=1,\cdots,n}$, \\
                            go to (b)(4); \\
                            else proceed to (b)(3)(ii).
                            \item [(ii)] If $l>1$, compute correlations inside batch: $c_{\mu j}^{(m-1)}=\rho(\boldsymbol{r}_{\mu}^{(m-1)}, \boldsymbol{g}_j)$, $j\in B_{\mu k}$.
                            \item [(iii)] Choose variant $j^*$ with the highest absolute correlation $|c_{\mu j^*}^{(m-1)}|=\max_{j\in  B_{\mu k}}|c_{\mu j}^{(m-1)}|$. \\
                            If $|c_{\mu j^*}^{(m-1)}|<c_{\text{stop},\mu}$, set $F_{\text{stop},\mu}=\texttt{TRUE}$, $\hat{\boldsymbol{\beta}}^{(m)}=\hat{\boldsymbol{\beta}}^{(m-1)}$, $\boldsymbol{r}_{\mu}^{(m)}=\left [ \frac{y_i-\boldsymbol{g}_i'\hat{\boldsymbol{\beta}}^{(m)}}{\exp(2\boldsymbol{g}_i'\hat{\boldsymbol{\gamma}}^{(m-1)})} \right ]_{i=1,\cdots,n}$, go to (b)(4); \\
                            else proceed to (b)(3)(iv).
                            \item [(iv)] Fit linear model: $E(\boldsymbol{r}_{\mu}^{(m-1)})=\hat{\beta_0}+\hat{\beta}_{j*}\cdot\boldsymbol{g}_{j*}$
                            \item [(v)] Update coefficients and residuals: \\
                            $\hat{\beta}_0^{(m)}=\hat{\beta}_0^{(m-1)}+\nu\cdot\hat{\beta}_0$, \\
                            $\hat{\beta}_{j^*}^{(m)}=\hat{\beta}_{j^*}^{(m-1)}+\nu\cdot\hat{\beta}_{j^*}$, \\
                            $\hat{\beta}_{j}^{(m)}=\hat{\beta}_{j}^{(m-1)},\, j\in\{1,\cdots, p\}\setminus \{j^*\}  $, \\
                            $\boldsymbol{r}_{\mu}^{(m)}=\left [ \frac{y_i-\boldsymbol{g}_i'\hat{\boldsymbol{\beta}}^{(m)}}{\exp(2\boldsymbol{g}_i'\hat{\boldsymbol{\gamma}}^{(m-1)})} \right ]_{i=1,\cdots,n}$.
                        \end{enumerate}

                        \item[(4)] For $\sigma$: \\
                        \begin{enumerate}
                            \item [(i)] If $F_{\text{stop},\sigma}=\texttt{TRUE}$, \\
                            $\hat{\boldsymbol{\gamma}}^{(m)}=\hat{\boldsymbol{\gamma}}^{(m-1)}$, $\boldsymbol{r}_{\sigma}^{(m)}=\left [ \frac{(y_i-\boldsymbol{g}_i'\hat{\boldsymbol{\beta}}^{(m)})^2}{\exp(2\boldsymbol{g}_i'\hat{\boldsymbol{\gamma}}^{(m)})}-1 \right ]_{i=1,\cdots,n}$, $l:=l+1$, \\
                            go to (b)(1); \\
                            else proceed to (b)(4)(ii).
                            \item [(ii)] If $l>1$, compute correlations inside batch: $c_{\sigma j}^{(m-1)}=\rho(\boldsymbol{r}_{\sigma}^{(m-1)}, \boldsymbol{g}_j)$, $j\in B_{\sigma k}$.
                            \item [(iii)] Choose variant $j^{\dagger}$ with the highest absolute correlation $|c_{\sigma j^{\dagger}}^{(m-1)}|=\max_{j\in  B_{\sigma k}}|c_{\sigma j}^{(m-1)}|$. \\
                            If $|c_{\sigma j^{\dagger}}^{(m-1)}|<c_{\text{stop},\sigma}$, set $F_{\text{stop},\sigma}=\texttt{TRUE}$, $\hat{\boldsymbol{\gamma}}^{(m)}=\hat{\boldsymbol{\gamma}}^{(m-1)}$, $\boldsymbol{r}_{\sigma}^{(m)}=\left [ \frac{(y_i-\boldsymbol{g}_i'\hat{\boldsymbol{\beta}}^{(m)})^2}{\exp(2\boldsymbol{g}_i'\hat{\boldsymbol{\gamma}}^{(m)})} - 1\right ]_{i=1,\cdots,n}$, $l:=l+1$,  go to (b)(1); \\
                            else proceed to (b)(4)(iv).  
                            \item [(iv)] Fit linear model: $E(\boldsymbol{r}_{\sigma}^{(m-1)})=\hat{\gamma_0}+\hat{\gamma}_{j^{\dagger}}\cdot\boldsymbol{g}_{j^{\dagger}}$
                            \item [(v)] Update coefficients and residuals: \\
                            $\hat{\gamma}_0^{(m)}=\hat{\gamma}_0^{(m-1)}+\nu\cdot\hat{\gamma}_0$, \\
                            $\hat{\gamma}_{j^{\dagger}}^{(m)}=\hat{\gamma}_{j^{\dagger}}^{(m-1)}+\nu\cdot\hat{\gamma}_{j^{\dagger}}$, \\
                            $\hat{\gamma}_{j}^{(m)}=\hat{\gamma}_{j}^{(m-1)},\, j\in\{1,\cdots, p\}\setminus \{j^{\dagger}\}  $, \\
                            $\boldsymbol{r}_{\sigma}^{(m)}=\left [ \frac{(y_i-\boldsymbol{g}_i'\hat{\boldsymbol{\beta}}^{(m)})^2}{\exp(2\boldsymbol{g_i}'\hat{\boldsymbol{\gamma}}^{(m)})} -1 \right ]_{i=1,\cdots,n}$.
                            \item [(vi)] $l:=l+1$, go to (b)(1).
                        \end{enumerate}

                    \end{enumerate}
                \end{enumerate}
            \end{enumerate}
        }   
        
\end{algorithm}

\newpage

\begin{algorithm}[!ht]
\caption*{\textbf{SNPBOOSTLSS (Continued)}}
\addcontentsline{loa}{algorithm}{My Algorithm}   
        \KwResult{
            \begin{enumerate} 
                \item[] 
                \begin{enumerate}
                    \item[(c)] If $k=b_{\text{max}}$ or if the loss function on the validation set has not decreased for $b_{\text{stop}}$ batches, end the outer loop; \\
                    else $k:=k+1$ and repeat (a)-(b).
                \end{enumerate}
                \item[3.] \textbf{Final model choice:} \\
                Find $m_{\text{stop}} \in \{1,\cdots,m\}$ corresponding to the lowest loss on validation set. The final coefficient estimates are given by $\hat{\boldsymbol{\beta}}^{(m_{\text{stop}})}$ and $\hat{\boldsymbol{\gamma}}^{(m_{\text{stop}})}$.
            \end{enumerate}
        }
\end{algorithm}

\newpage

\section{Additional simulation studies}

\subsection{Comparison between fixed step length and adaptive step length}
\label{sec:ASL_FSL}

Traditional gradient boosting often uses fixed step-lengths for updating the model coefficients, regardless of the achieved loss reduction for different distribution parameters. But different parameters affect the magnitude of loss differently, and an update of the same size on all predictors hence results in different improvements with respect to loss reduction. This may lead to imbalanced updates that affect the fair selection between parameters. Zhang et al. (2022) proprosed using instead adaptive step lengths for Gaussian location-scale model to balance the updates between parameters. In the $m$-th iteration of boosting update, the adaptive lengths for mean and variance parameters are given as follows:
\begin{equation}\label{eqn:asl}
\nu_{j^*,\mu}^{(m)}=\lambda\cdot\frac{\sum^n_{i=1}(\hat{h}_{j^*,\mu}(g_{ij^*}))^2}{\sum^n_{i=1}\frac{(\hat{h}_{j^*,\mu}(g_{ij^*}))^2}{\hat{\sigma}_i^{2(m-1)}}}, \quad \nu_{j^*,\sigma}^{(m)}=0.05
\end{equation}
where $\hat{h}_{j^*,\mu}(g_{ij^*})=\hat{\beta}_0+\hat{\beta}_{j^*}\cdot g_{ij^*}$ is the fitted base learner in $m$-th iteration for mean and $\hat{\sigma}_i^{2(m-1)}$ is the estimated variance after $m-1$ iterations. $\lambda$ is a shrinkage parameter with a suggested default value of 0.1. Regarding the adaptive step length for updating variance parameter $\sigma$, Zhang et al. (2022) found that the optimal step length is in general hard to calculate as there is no closed-form solution and its limiting value of 0.05 can already provide a good approximation and yield satisfactory performance. Therefore, we also take 0.05 as the default step length for $\sigma$ under adaptive step length option.

\begin{figure}[ht]
\centering
\includegraphics[width=\linewidth]{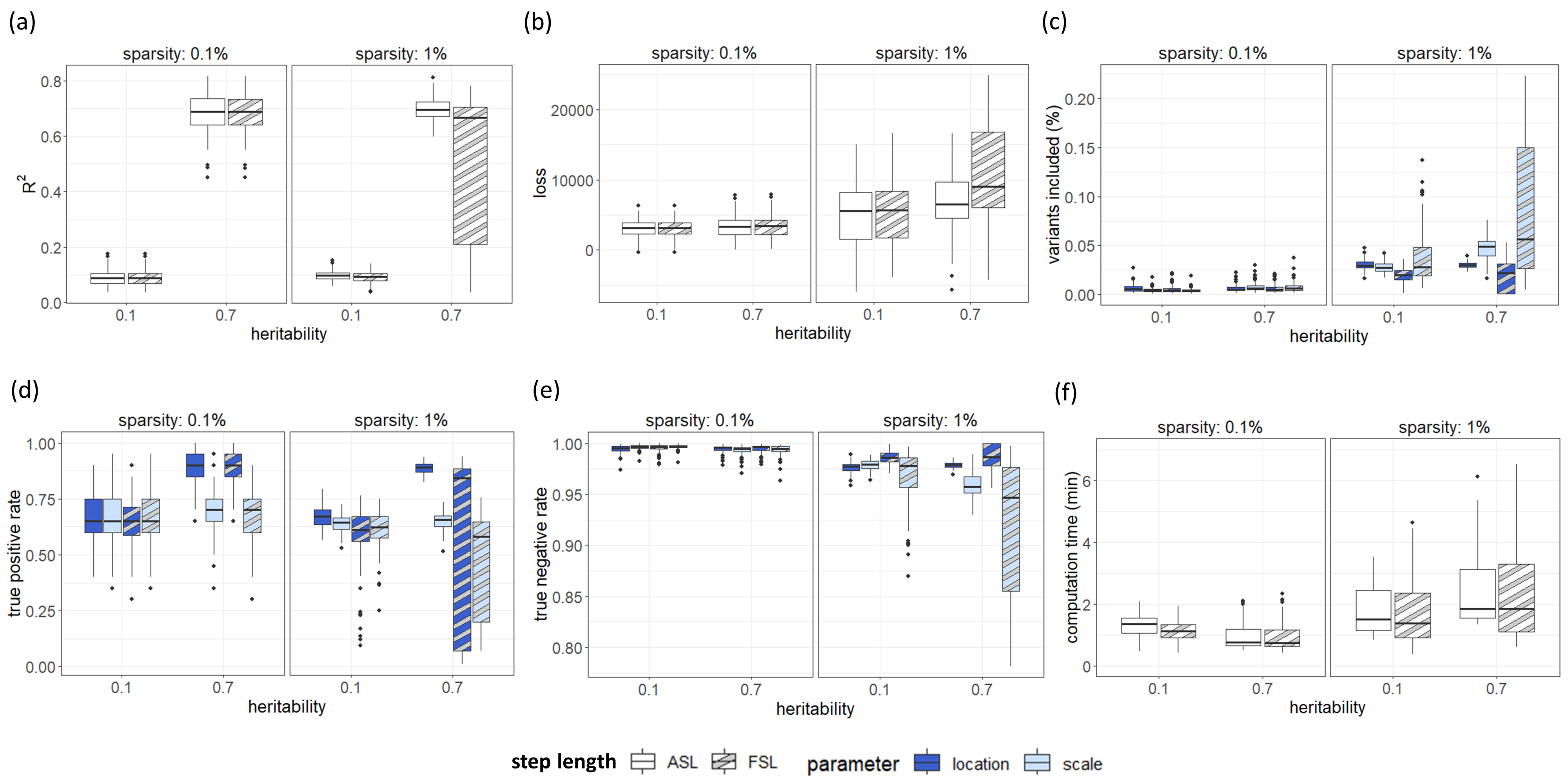}
\caption{Comparison between fixed step length and adaptive step length. Results of scenarios with heritability $h^2\in\{0.1,0.7\}$ and sparsity $s\in\{0.1\%,1\%\}$ for $p=20,000$ variants and $n=20,000$ individuals (divided into 50\% training, 20\% validation and 30\% test sets) are shown. For each performance metric, the boxplots from 100 simulations are displayed.}
\label{fig:sim_res_asl_fsl}
\end{figure}

We conducted a simulation study to compare the effect of using adaptive step lengths (ASL) in \eqref{eqn:asl} versus traditional fixed step length (FSL) of value 0.1 in snpboostlss. The data generating mechanism and performance measures are the same as described in \emph{Methods, Simulation settings}. Results are shown in the Figure~\ref{fig:sim_res_asl_fsl}.

\begin{figure}[ht]
\centering
\includegraphics[width=\linewidth]{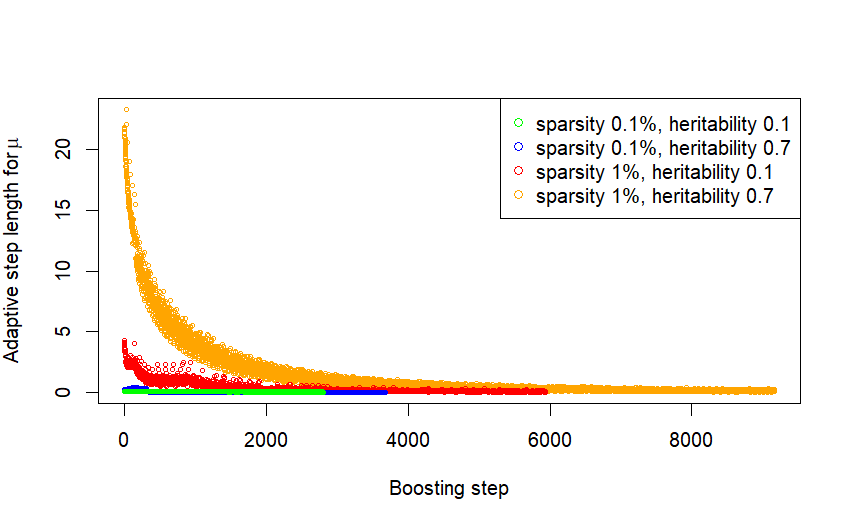}
\caption{Changes in adaptive step length for $\mu$ over boosting iterations. One simulation from each scenario is randomly selected as examples for illustration.}
\label{fig:asl_example}
\end{figure}

Figure~\ref{fig:sim_res_asl_fsl}(a) and (b) show that ASL achieves similar prediction performance as FSL when number of informative variants is low (i.e., 0.1\% sparsity setting), while outperforms FSL when more variants are informative (i.e., 1\% sparsity setting). Its prediction performance is also much stabler than FSL at 1\% sparsity level. The motivation to consider ASL is to achieve balanced updates between $\mu$ and $\sigma$. This is verified in Figure~\ref{fig:sim_res_asl_fsl}(c) especially when sparsity level is 1\%. The number of informative variants is the same for both parameters, but the average number of variants selected for two parameters are much more divergent using FSL than using ASL. In terms of variable selection accuracy, ASL in general achieves higher true positive rate and true negative rate than FSL (Figure~\ref{fig:sim_res_asl_fsl}(d) and (e)). In addition, ASL and FSL take similar computation time but FSL yields more volatility in computation time (Figure~\ref{fig:sim_res_asl_fsl}(f)). An illustration of the adaptive step lengths for updating $\mu$ in 4 randomly selected simulation runs can be found in Figure~\ref{fig:asl_example}. In summary, ASL, in comparison to FSL, achieves better prediction performance, more balanced updates between parameters and higher variable selection accuracy. Such advantages are more prominent when there are many informative variants with large effect size. Therefore, we set ASL as the default step length for snpboostlss.

\subsection{Sanity check on the performance of snpboostlss}

\begin{figure}[ht]
\centering
\includegraphics[width=\linewidth]{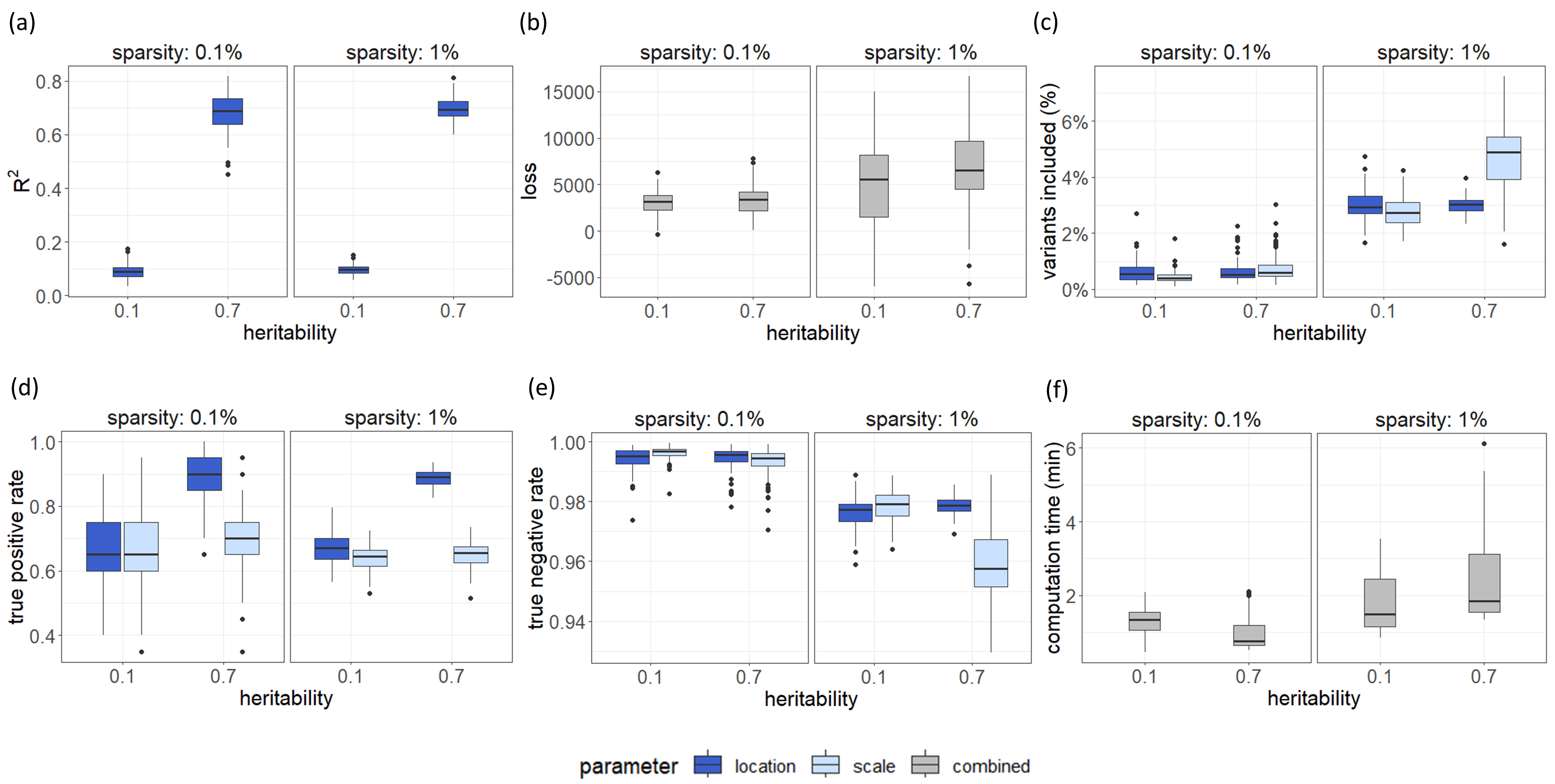}
\caption{Performance of snpboostlss. Results of scenarios with heritability $h^2\in\{0.1,0.7\}$ and sparsity $s\in\{0.1\%,1\%\}$ for $p=20,000$ variants and $n=20,000$ individuals (divided into 50\% training, 20\% validation and 30\% test sets) are shown. For each performance metric, the boxplots from 100 simulations are displayed.}
\label{fig:sim_res_aim1_main}
\end{figure}

We conducted a simulation study to check the performance of snpboostlss with default parameter values. Figure~\ref{fig:sim_res_aim1_main} shows the performance of snpboostlss in terms of prediction accuracy, selection of informative variants and computation time. The $R^2$ values in Figure~\ref{fig:sim_res_aim1_main}(a) are very close to the true heritability in each scenario, indicating an accurate capture of genetic susceptibility for the phenotypic mean. When evaluating the prediction performance via loss defined as negative log-likelihood (Figure~\ref{fig:sim_res_aim1_main}(b)), which takes both mPRS and vPRS into account, the loss becomes larger and more volatile when the proportion of informative variants increases from 0.1\% to 1\%, because more complex models increase the difficulty of model fitting.  Figure~\ref{fig:sim_res_aim1_main}(c) reflects the common phenomenon that boosting has the tendency to overestimate the number of informative variants. With adaptive step length, we are able to achieve balanced updates between two PRS models, namely similar number of variants are included in mPRS and vPRS for most scenarios (Section \ref{sec:ASL_FSL}). The only exception is when heritability is high and there are more informative variants, which creates more challenges for modeling vPRS because of the difficulty in this case to detect the weak signal for $\sigma$. Regarding variant selection accuracy, snpboostlss achieves a satisfactory average true positive rate for both PRS models in all scenarios and performs particularly well on mPRS when heritability is high (Figure~\ref{fig:sim_res_aim1_main}(d)). In terms of true negative rate, more than 95\% of non-informative variants are correctly excluded from mPRS or vPRS in all scenarios (Figure~\ref{fig:sim_res_aim1_main}(e)). Despite the complexity of the model and the challenging data situation, most simulation runs take less than three minutes. As expected, computation time increases when there are more informative variants to be estimated (Figure~\ref{fig:sim_res_aim1_main}(f)). To summarize, we investigated the performance of snpboostlss under different genetic architectures by considering different combinations of heritability and sparsity. We found that under different simulation settings the prediction performances for mPRS scales with the heritability and therefore snpboostlss can properly model the genetic liability underlying polygenic traits. The algorithm achieves balanced updates between PRS models and make accurate inclusion/exclusion decisions for most variants in an efficient manner. 

\newpage

\section{Real data application on UK Biobank}

\subsection{Analysis flowchart}

\begin{figure}[!h]
\centering
\includegraphics[scale=0.6]{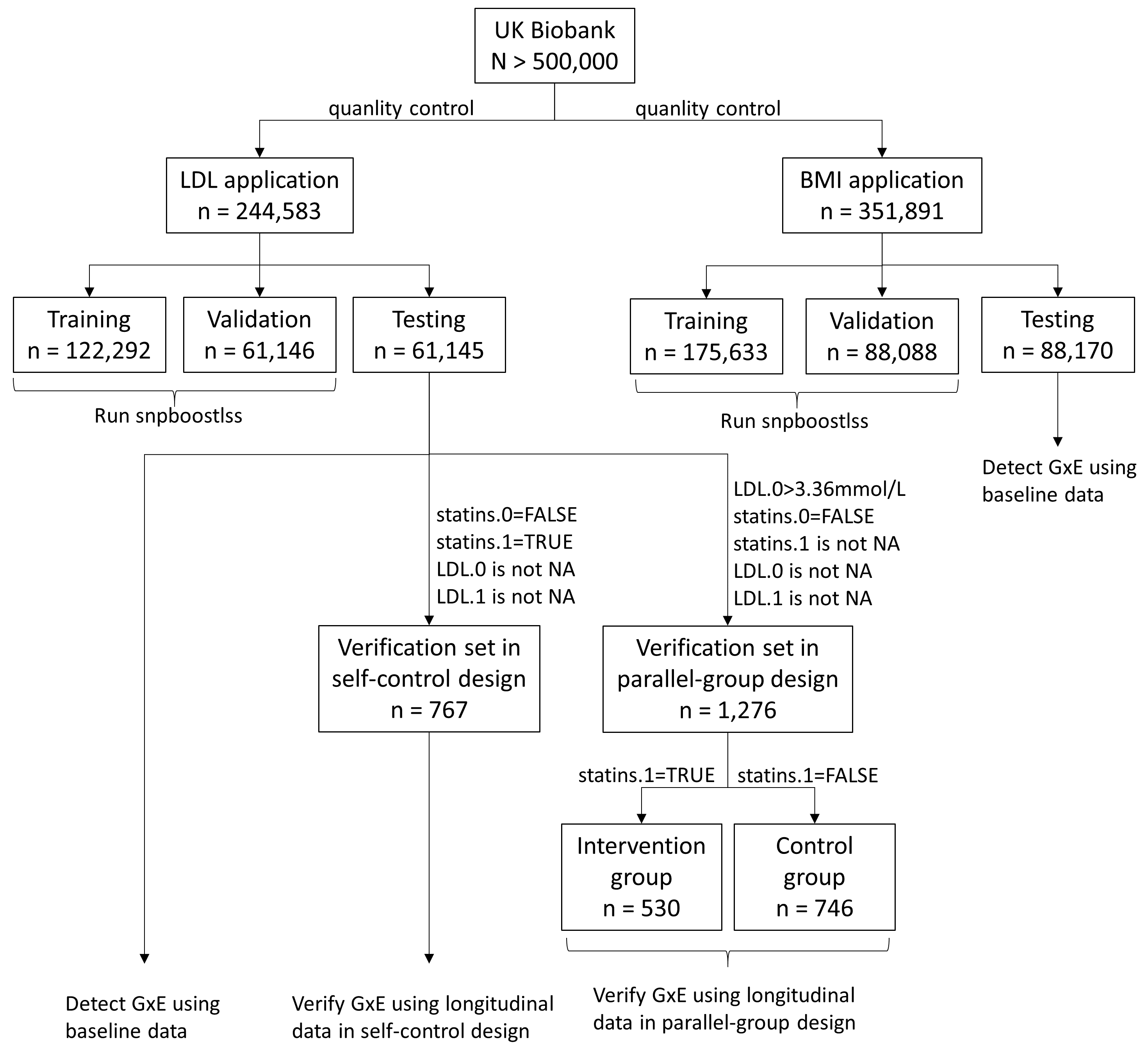}
\caption{Analysis flowchart for the real data application on UK Biobank. LDL.0 and LDL.1 are measurements of LDL at baseline and first revisit. Statins.0 and statins.1 are the usage status of statins at baseline and first revisit.}
\label{fig:application_flowchart}
\end{figure}

\newpage

\subsection{LDL application}

\subsubsection{Distribution of LDL in UK Biobank}

\begin{figure}[!h]
\centering
\includegraphics[width=\linewidth]{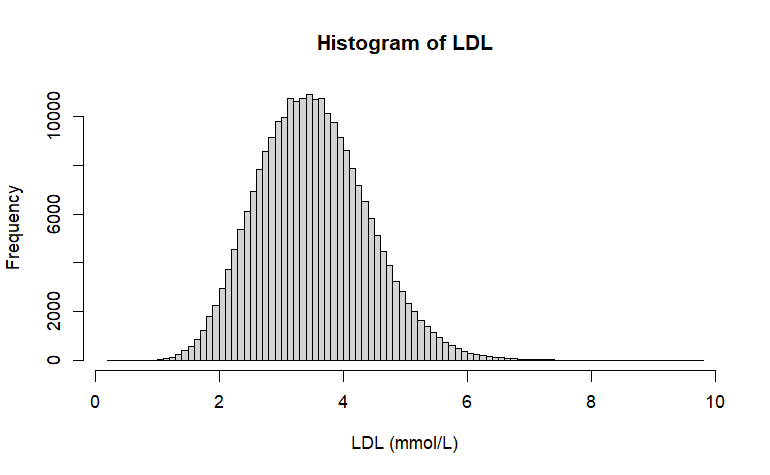}
\caption{Histogram of LDL on $224,583$ subjects from UK Biobank.}
\label{fig:ldl_hist}
\end{figure}

\subsubsection{LDL: Detection of GxE using baseline data}

\begin{table}[ht]
\centering
\caption{Estimated vPRSxE effects on LDL in UK Biobank}
\begin{tabular}{lllll}
\toprule
\multicolumn{1}{l}{\textbf{Basic analysis}}  &       & \multicolumn{1}{l}{} &              &  \\ \hline
\multicolumn{1}{l|}{Environmental factor} & Main effect & \multicolumn{1}{l|}{P-value}          & Interaction effect & P-value  \\ \hline
\multicolumn{1}{l|}{statins usage status}               & -1.106      & \multicolumn{1}{l|}{$<2\times 10^{-16}$} & -0.088             & $<2\times 10^{-16}$ \\ \hline
\multicolumn{1}{l}{\textbf{Robust analysis}}  &       & \multicolumn{1}{l}{} &              &  \\ \hline
\multicolumn{1}{l|}{Environmental factor} & Main effect & \multicolumn{1}{l|}{P-value}          & Interaction effect & P-value  \\ \hline
\multicolumn{1}{l|}{statins usage status}               & -1.106      & \multicolumn{1}{l|}{$<2\times 10^{-16}$} & -0.074             & $<2\times 10^{-16}$ \\ \bottomrule
\end{tabular}
\end{table}
\vspace{-0.15in}
\footnotesize{Basic analysis model: $\text{Y}\sim \text{mPRS}+\text{vPRS}+\text{E}+\text{vPRS}\times \text{E}+\text{age}+\text{sex}+\text{(genotying array)}+\text{(top 12 PCs)}$. Robust analysis model adds two additional interaction terms: $\text{vPRS}\times \text{age}$ and $\text{vPRS}\times \text{sex}$.}

\subsubsection{LDL: Verification of GxE using longitudinal data in parallel group design}

\begin{figure}[H]
\centering
\includegraphics[scale=0.47]{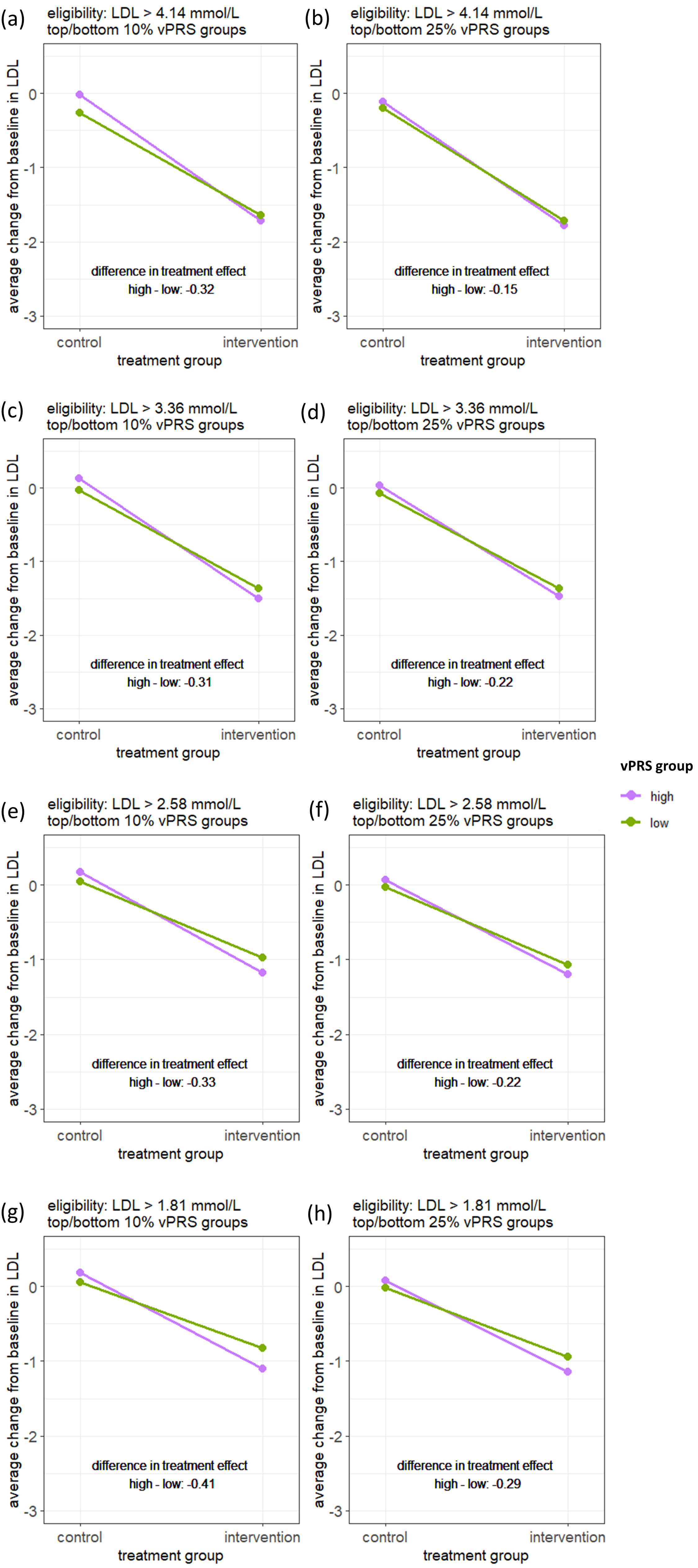}
\caption{Comparison of statins treatment effect between high- and low-vPRS groups in parallel group design. Different LDL thresholds for eligibility criteria and different high/low vPRS subgrouping are considered.}
\label{fig:more_thresholds_tendency}
\end{figure}

\begin{figure}[!h]
\centering
\includegraphics[width=\linewidth]{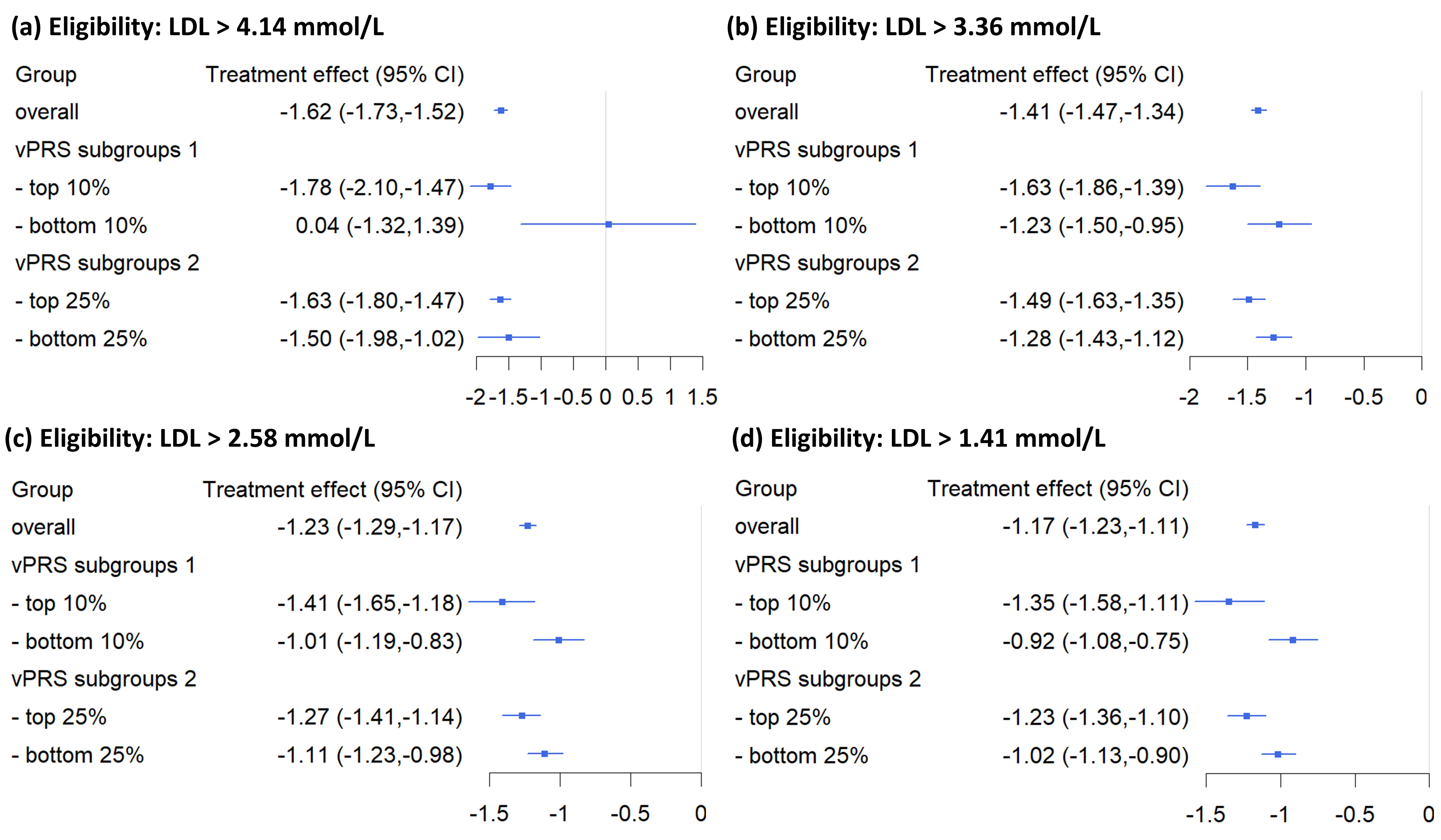}
\caption{Overall and vPRS-subgroup treatment effect of statins in parallel group design. The overall analysis set was obtained by screening the test set based on different eligibility criteria: (a) LDL > 4.14 mmol/L, (b) LDL > 3.36 mmol/L, (c) LDL > 2.58 mmol/L or (d) LDL > 1.81 mmol/L. High/low vPRS groups are defined as people with vPRS beyong 90\%/10\% or 75\%/25\% percentile of vPRS in test set. Treatment effect is obtained from linear model with LDL change from baseline as response and adjusted for treatment group and other baseline covariates.}
\label{fig:more_threshold_forest}
\end{figure}

\newpage

\subsection{BMI application}

\subsubsection{Distribution of BMI in UK Biobank}

\begin{figure}[H]
\centering
\includegraphics[width=\linewidth]{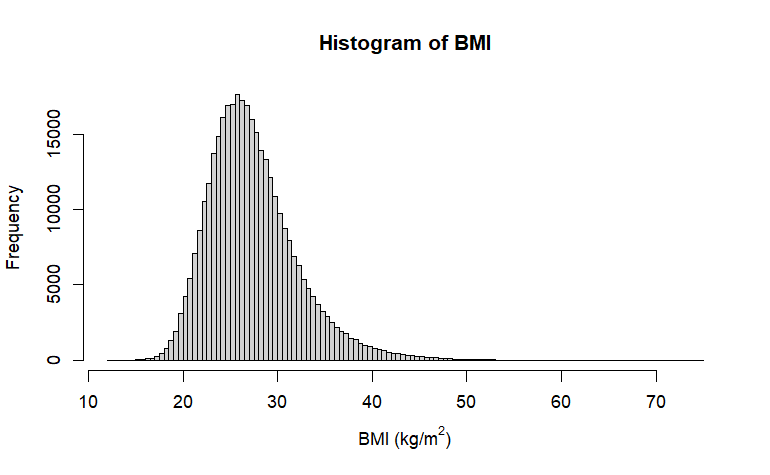}
\caption{Histogram of BMI on $351,891$ subjects from UK Biobank.}
\label{fig:bmi_hist}
\end{figure}

\subsubsection{BMI: Detection of GxE using baseline data}

\begin{table}[!h]
\centering
\caption{Estimated vPRSxE effects on BMI in UK Biobank}
\begin{tabular}{lllll}
\toprule

\multicolumn{1}{l}{\textbf{Basic analysis}}  &       & \multicolumn{1}{l}{} &              &  \\ \hline
\multicolumn{1}{l|}{Environmental factor} & Main effect & \multicolumn{1}{l|}{P-value}          & Interaction effect & P-value  \\ \hline
\multicolumn{1}{l|}{physical activity}               & -0.760      & \multicolumn{1}{l|}{$<2\times 10^{-16}$} & -0.066             & $8.73\times 10^{-4}$ \\
\multicolumn{1}{l|}{sedentary behavior}               & 0.422       & \multicolumn{1}{l|}{$<2\times 10^{-16}$} & 0.020              & $1.31\times 10^{-3}$ \\ \hline
\multicolumn{1}{l}{\textbf{Robust analysis}}  &       & \multicolumn{1}{l}{} &              &  \\ \hline
\multicolumn{1}{l|}{Environmental factor} & Main effect & \multicolumn{1}{l|}{P-value}          & Interaction effect & P-value  \\ \hline
\multicolumn{1}{l|}{physical activity}               & -0.760      & \multicolumn{1}{l|}{$<2\times 10^{-16}$} & -0.063             & $1.33\times 10^{-3}$  \\
\multicolumn{1}{l|}{sedentary behavior}               & 0.422       & \multicolumn{1}{l|}{$<2\times 10^{-16}$} & 0.027              & $2.72\times 10^{-5}$ \\ \bottomrule
\end{tabular}
\end{table}
\vspace{-0.15in}
\footnotesize{Basic analysis model: $\text{Y}\sim \text{mPRS}+\text{vPRS}+\text{E}+\text{vPRS}\times \text{E}+\text{age}+\text{sex}+\text{(genotying array)}+\text{(top 12 PCs)}$. Robust analysis model adds two additional interaction terms: $\text{vPRS}\times \text{age}$ and $\text{vPRS}\times \text{sex}$.}
